\documentclass[12pt]{article}   
\usepackage{epsfig}
\newcommand{\mysection}{\setcounter{equation}{0}\section}
\renewcommand{\theequation}{\thesection.\arabic{equation}}
\def\beq{\begin{equation}}
\def\eeq{\end{equation}}
\def\beqa{\begin{eqnarray}}
\def\eeqa{\end{eqnarray}}

\newlength{\dinwidth} \newlength{\dinmargin}
\begin{document}

\begin{center}
{\Large \bf Next-to-next-to-next-to-leading-order soft-gluon corrections 
in hard-scattering processes near threshold}
\end{center}
\vspace{2mm}
\begin{center}
{\large Nikolaos Kidonakis}\\
\vspace{2mm}
{\it Kennesaw State University, Physics \#1202\\
1000 Chastain Rd., Kennesaw, GA 30144-5591}
\end{center}

\begin{abstract}
I present a unified calculation of 
soft-gluon corrections to hard-scattering cross sections 
through next-to-next-to-next-to-leading order (NNNLO). 
Master formulas are derived, from a threshold resummation 
formalism, that can be applied to total and differential cross sections 
for hard-scattering processes in hadron colliders.
I also present numerical results for charged Higgs 
production at the LHC where these corrections are large, and for 
top quark production at the Tevatron where these corrections greatly reduce
the scale dependence of the cross section.  
\end{abstract}

\thispagestyle{empty} \newpage \setcounter{page}{2}

\mysection{Introduction}

Calculations of hard-scattering cross sections become very complicated
as one moves from the lowest order to higher-order corrections.
Current theoretical approaches include a variety of resummations
and fixed-order calculations, some through next-to-next-to-leading order 
(NNLO) \cite{QCDSM}. These tools can greatly improve
next-to-leading order (NLO) calculations because
higher-order corrections reduce the scale dependence and increase 
theoretical accuracy.

The QCD corrections can be separated  into hard, soft,
and virtual parts, corresponding to contributions from energetic, soft, 
and virtual gluons, respectively.
The soft-gluon corrections are an important component of  
the total result and, in some
schemes and kinematical regions, e.g. threshold, they are 
numerically dominant.
There is a universality in the form of these soft-gluon 
corrections, as can be clearly seen from the techniques of threshold 
resummations, which formally resum the soft-gluon contributions to all 
orders in the strong coupling \cite{GS,CT,CLS,KS,KOS,LOS,NK3nlo}. 
Although resummed calculations are 
prescription-dependent (see discussion in Ref. \cite{NKtop}),
finite-order expansions of resummed cross sections are not, and they 
have provided us with many cross sections with NNLO soft-gluon corrections 
(for a review see Ref. \cite{NKmpla}).
Examples include $W$-boson production at large transverse momentum \cite{GKS},
direct photon production \cite{KOp}, top quark production 
\cite{NKtop,KLMV,NKRVtop}, bottom and charm quark
production \cite{KVbc}, heavy quark electroproduction \cite{LM}
and photoproduction \cite{NYI}, 
charged Higgs production \cite{NKchiggs}, 
and jet production \cite{KOj}.

The NLO soft-gluon corrections are typically a very good approximation to 
the exact NLO corrections near threshold.
The NNLO soft-gluon corrections can be numerically significant 
and they invariably improve the
theoretical calculation by stabilizing the dependence of the cross
section on the factorization and renormalization scales, which are
arbitrary energy scales in the theory.
Thus it is worthwhile to provide a unified approach for the calculation 
of these and even higher-order soft-gluon corrections
to hard-scattering processes in hadron colliders. 
It is important to note that new particles, such as in supersymmetry 
\cite{SUSY} or Higgs physics \cite{Higgs}, 
or particle production via new interactions, 
such as top production via flavor-changing neutral currents \cite{BK}, 
will likely be discovered
near threshold where the soft-gluon corrections are dominant, therefore  
soft-gluon calculations are relevant to more than just pure QCD processes.

The calculation of hard-scattering cross sections in hadron-hadron or 
lepton-hadron collisions can be written as
\beqa
\sigma=\sum_f \int \left[ \prod_i  dx_i \, \phi_{f/h_i}(x_i,\mu_F)\right]\,
{\hat \sigma}(s,t_i,\mu_F,\mu_R) \, ,
\label{factcs}
\eeqa
where $\sigma$ is the physical cross section,
$\phi_{f/h_i}$ is the distribution function for parton $f$ carrying momentum 
fraction $x_i$ of hadron $h_i$, at a factorization scale $\mu_F$, 
and $\mu_R$ is the renormalization scale.
The parton-level cross section is denoted by 
${\hat \sigma}$, and $s$, $t_i$ are standard kinematical 
invariants formed from the 4-momenta of the particles in the hard scattering.
In a lepton-hadron collision we have one parton distribution
($i=1$) while in a hadron-hadron collision we have two, $i=1,2$.
The partonic processes are of the form
\beq
f_{1}(p_1)\, + \, f_{2}\,[l_2](p_2) \rightarrow F\, + \, X \, ,
\eeq
where $f$ [$l$] represents a parton [lepton],
$F$ represents an observed system in the final state, and $X$ any additional
allowed final-state particles. For example, 
$F$ can represent a pair of heavy quarks, a single heavy quark, 
a jet, a photon, a Higgs boson, a pair of squarks, etc.
Then $s=(p_1+p_2)^2$. In single-particle-inclusive (1PI) kinematics 
we identify one particle $F$ with momentum $p$, and define 
the kinematical invariants $t_1=(p_1-p)^2$, $t_2=(p_2-p)^2$ 
(also commonly denoted by $t$ and $u$, respectively).
In pair-invariant-mass (PIM) kinematics we identify a pair of particles
(such as a heavy quark antiquark pair) with invariant mass squared $Q^2$. 
We note here that $\sigma$ and ${\hat \sigma}$ are not restricted to be
total cross sections; they can represent any differential cross 
section of interest.

In general, $\hat{\sigma}$ includes plus distributions ${\cal D}_l(x_{th})$ 
with respect to a kinematical variable 
$x_{th}$ that measures distance from threshold, with $l \le 2n-1$ at $n$th 
order in $\alpha_s$ beyond the leading order. These are the soft  
corrections. 
The virtual corrections multiply delta functions $\delta(x_{th})$.
In 1PI kinematics, 
$x_{th}$ is usually denoted as $s_4$
(or $s_2$), $s_4=s+t_1+t_2-\sum m^2$ (the sum is over the masses 
squared of all particles in the process), and it vanishes at threshold. 
The plus distributions are of the form 
\beq
{\cal D}_l(s_4)\equiv\left[\frac{\ln^l(s_4/M^2)}{s_4}\right]_+ \, ,
\eeq
where $M^2$ is a hard scale relevant to the process at hand, for example
the mass $m$ of a heavy quark, the transverse momentum $p_T$ of a jet, etc.
The distributions  are defined through their integral with any smooth 
function, such as parton densities, by 
\beqa
\int_0^{s_{4 \, max}} ds_4 \, f(s_4) \left[\frac{\ln^l(s_4/M^2)}
{s_4}\right]_{+} &\equiv&
\int_0^{s_{4\, max}} ds_4 \frac{\ln^l(s_4/M^2)}{s_4} [f(s_4) - f(0)]
\nonumber \\ &&
{}+\frac{1}{l+1} \ln^{l+1}\left(\frac{s_{4\, max}}{M^2}\right) f(0) \, .
\label{1piplus}
\eeqa
In PIM kinematics, with $Q^2$ the invariant mass 
squared of the produced pair, $x_{th}$ is usually called $1-x$ or $1-z$, 
with $z=Q^2/s \rightarrow 1$ at threshold.  Then the 
plus distributions are of the form
\beq
{\cal D}_l(z)\equiv\left[\frac{\ln^l(1-z)}{1-z}\right]_+
\eeq
defined through their integral with any smooth function by
\beqa
\int_y^1 dz \, f(z) \left[\frac{\ln^l(1-z)}{1-z}\right]_{+} &\equiv&
\int_y^1 dz \frac{\ln^l(1-z)}{1-z} [f(z) - f(1)]
\nonumber \\ &&
{}+\frac{1}{l+1} \ln^{l+1}(y) f(1) \, .
\label{pimplus}
\eeqa
The highest powers of these distributions in the $n$th-order corrections
are the leading logarithms (LL) with $l=2n-1$,
the second highest are the next-to-leading logarithms (NLL) with $l=2n-2$, 
the third highest are the next-to-next-to-leading logarithms (NNLL) with 
$l=2n-3$,  the fourth highest are the  next-to-next-to-next-to-leading 
logarithms (NNNLL) with $l=2n-4$, etc. These logarithms can be in 
principle resummed to all orders in perturbation theory.
By now there are several processes for which NLL resummations and
NNLO-NNLL results (i.e. the NNLL terms\footnote{The counting of logarithms 
is different in the exponent and in the fixed-order expansions; for example
a term that is NNLL in the fixed-order expansion, as described here, 
may be NLL in the exponent \cite{KLMV}.} at NNLO) have been 
presented \cite{NKmpla,NKuni}. 

In this paper I present master formulas for the NLO, NNLO, and NNNLO 
soft-gluon corrections for processes in hadron-hadron
or hadron-lepton collisions. 
These processes can be of QCD, electroweak, Higgs, or supersymmetric 
origin at lowest order.
Results on the NLO and NNLO corrections have
been presented before \cite{NKuni} 
but the notation here is somewhat different to
facilitate the calculation of the NNNLO corrections.
The NNNLO results and their applications to top quark and charged Higgs
production are new.
In the next section, I present a threshold resummation formula 
from which high-order expansions are derived. 
In Sections 3 and 4 are presented  master formulas for the NLO and NNLO soft  
corrections, respectively,  that arise from the expansion of the resummation
formula. The formulas are given in the $\overline{\rm MS}$
scheme (see \cite{NKuni} for results through NNLO in the DIS scheme), 
and cover both 1PI and PIM kinematics. 
In Section 5, I present a master formula for the NNNLO corrections.
In Sections 6 and 7 applications to charged Higgs production at the LHC
and top quark pair production at the Tevatron, respectively, are discussed.
Conclusions are given in Section 8. 
Some long expressions for terms in the NNNLO master formula 
are collected in an Appendix.

\mysection{Soft-gluon corrections from threshold resummation}

We begin with a brief review of the threshold resummation formalism.
Threshold resummation follows from factorization theorems for 
hard-scattering cross sections. One can write the hadronic cross section 
as a convolution of parton densities with a parton-level cross section.
This can be further refactorized into functions associated with soft and 
collinear gluon emission from the incoming partons and any outgoing 
partons or jets, a function associated with noncollinear soft gluon emission 
that involves the color structure of the hard scattering, and a short-distance
hard-scattering function. The renormalization group properties of these 
functions result in the exponentiation of the soft-gluon contributions 
thus providing the resummed cross section \cite{KS,KOS,LOS}. 
For a review see Ref. \cite{NKrev}.

The resummation of threshold logarithms is carried out in moment space.
We define moments of the partonic cross section by
${\hat\sigma}(N)=\int dz \, z^{N-1} {\hat\sigma}(z)$ (PIM) or by  
${\hat\sigma}(N)=\int (ds_4/s) \;  e^{-N s_4/s} {\hat\sigma}(s_4)$ (1PI),
with $N$ the moment variable. The logarithms of $N$ exponentiate.
The resummed partonic cross section in moment space is then given by 
\beqa
{\hat{\sigma}}^{res}(N) &=&   
\exp\left[ \sum_i E^{f_i}(N_i)\right] \; 
\exp\left[ \sum_j {E'}^{f_j}(N_j)\right] 
\nonumber\\ && \hspace{-10mm} \times \,
\exp \left[\sum_i 2\int_{\mu_F}^{\sqrt{s}} \frac{d\mu}{\mu}\;
\gamma_{i/i}\left(\alpha_s(\mu)\right)\right] \;
\exp\left[2\, d_{\alpha_s} \int_{\mu_R}^{\sqrt{s}}\frac{d\mu}{\mu}\; 
\beta\left(\alpha_s(\mu)\right)\right] 
\nonumber\\ && \hspace{-10mm} \times \,
{\rm Tr} \left \{H^{f_i f_j}\left(\alpha_s(\mu_R)\right) \;
\exp \left[\int_{\sqrt{s}}^{{\sqrt{s}}/{\tilde N_j}} 
\frac{d\mu}{\mu} \;
\Gamma_S^{\dagger \, f_i f_j}\left(\alpha_s(\mu)\right)\right] \;
{\tilde S^{f_i f_j}} \left(\alpha_s(\sqrt{s}/{\tilde N_j}) \right) \right. 
\nonumber\\ && \quad \left.\times \,
\exp \left[\int_{\sqrt{s}}^{{\sqrt{s}}/{\tilde N_j}} 
\frac{d\mu}{\mu}\; \Gamma_S^{f_i f_j}
\left(\alpha_s(\mu)\right)\right] \right\} \, .
\label{resHS}
\eeqa
The sums over $i$ run over incoming partons: in hadron-hadron
colisions we have two partons in the initial state, so $i=1,2$; 
in lepton-hadron collisions we have one parton. The sum over $j$ is relevant
if we have massless partons in the final state at lowest order. 
We note that we have suppressed all gauge-dependent terms 
because these terms cancel out explicitly.
  
Equation (\ref{resHS}) is actually valid for both 1PI and PIM kinematics
with appropriate definitions for $N_i$ and $N_j$.
In 1PI kinematics $N_i=N (-t_i/M^2)$ for incoming partons $i$,
and  $N_j=N (s/M^2)$ for outgoing partons $j$; here $M^2$ is any
chosen hard scale relevant to the process at hand.
In PIM kinematics $N_i=N_j=N$. 
Also note that ${\tilde N}=N e^{\gamma_E}$, with $\gamma_E$ the Euler constant.
The various exponents above are known at most to three loops.
Below we give explicitly only the one-loop (and some two-loop) expressions
that will be needed in our applications to charged Higgs production 
at NLL accuracy and top quark production at NNLL accuracy. 
Some two-loop and three-loop results 
can be found explicitly in \cite{NKuni} and \cite{CB,MVV}.
We note however that the two-loop and higher-loop results 
for process-dependent functions, such as the soft anomalous dimensions 
$\Gamma_S$, have to be calculated explicitly for a specified partonic process. 

The first exponent in Eq. (\ref{resHS}) is given in the $\overline{\rm MS}$ 
scheme by 
\beq
E^{f_i}(N_i)=
-\int^1_0 dz \frac{z^{N_i-1}-1}{1-z}\;
\left \{\int^1_{(1-z)^2} \frac{d\lambda}{\lambda}
A_i\left(\alpha_s(\lambda s)\right)
+{\nu}_i\left[\alpha_s((1-z)^2 s)\right]\right\} \, ,
\label{Eexp}
\eeq
with $A_i(\alpha_s) = A_i^{(1)} {\alpha_s/\pi}+A_i^{(2)} ({\alpha_s/\pi})^2
+A_i^{(3)}({\alpha_s/\pi})^3+\cdots$.
Here $A_i^{(1)}=C_i$ with $C_i=C_F=(N_c^2-1)/(2N_c)$ for a quark 
or antiquark and $C_i=C_A=N_c$ for a gluon, with $N_c$ the number of colors,
while $A_i^{(2)}=C_i K/2$ with $K= C_A\; ( 67/18-\pi^2/6 ) - 5n_f/9$ \cite{KT},
where $n_f$ is the number of quark flavors. 
Also ${\nu}_i=(\alpha_s/\pi){\nu}_i^{(1)}+(\alpha_s/\pi)^2 {\nu}_i^{(2)}
+(\alpha_s/\pi)^3 {\nu}_i^{(3)}+\cdots$, with ${\nu}_i^{(1)}=C_i$. 

The second exponent is given by 
\beq
{E'}^{f_j}(N_j)=
\int^1_0 dz \frac{z^{N_j-1}-1}{1-z}\;
\left \{\int^{1-z}_{(1-z)^2} \frac{d\lambda}{\lambda}
A_j \left(\alpha_s\left(\lambda s\right)\right)
-B_j\left[\alpha_s((1-z)s)\right]
-\nu_{j}\left[\alpha_s((1-z)^2 s)\right]\right\} \, ,
\label{Ejexp}
\eeq
where $B_j=(\alpha_s/\pi)B_j^{(1)}+(\alpha_s/\pi)^2 B_j^{(2)}
+(\alpha_s/\pi)^3 B_j^{(3)}+\cdots$
with $B_q^{(1)}=3C_F/4$ and $B_g^{(1)}=\beta_0/4$, where
$\beta_0$ is the lowest-order $\beta$-function, $\beta_0=(11C_A-2n_f)/3$.

In the third exponent $\gamma_{i/i}$ is the moment-space 
anomalous dimension of the ${\overline {\rm MS}}$ density $\phi_{i/i}$.
Note that the $N$-independent part of the one-loop ${\gamma}_{i/i}$ 
is the same as $\gamma_i^{(1)}$, the one-loop parton anomalous 
dimension, given by  $\gamma_q^{(1)}=3C_F/4$ and $\gamma_g^{(1)}=\beta_0/4$ 
for quarks and gluons, respectively \cite{KLMV}. 

The $\beta$ function in the fourth exponent is given in the Appendix.
The constant $d_{\alpha_s}=0,1,2$ if the Born cross section is of order
$\alpha_s^0$, $\alpha_s^1$, $\alpha_s^2$, respectively.

$H^{f_if_j}$ are the hard-scattering functions 
for the scattering of partons $f_i$ and $f_j$, while $S^{f_if_j}$ are the 
soft functions describing noncollinear soft gluon emission.
We use the expansions 
$H=\alpha_s^{d_{\alpha_s}}H^{(0)}+(\alpha_s^{d_{\alpha_s}+1}/\pi)H^{(1)}
+(\alpha_s^{d_{\alpha_s}+2}/\pi^2)H^{(2)}
+(\alpha_s^{d_{\alpha_s}+3}/\pi^3)H^{(3)}
+\cdots$ and
$S=S^{(0)}+(\alpha_s/\pi)S^{(1)}+(\alpha_s/\pi)^2 S^{(2)}
+(\alpha_s/\pi)^3 S^{(3)}+\cdots$. 
Note that both $H$ and $S$ are matrices in color space and 
the trace is taken. 
At lowest order, the trace of the product of the hard matrices
$H$ and soft matrices $S$
reproduces the Born cross section for each partonic process,
$\sigma^B=\alpha_s^{d_{\alpha_s}}{\rm tr}[H^{(0)}S^{(0)}]$.
The evolution of the soft function  
follows from its renormalization group properties and 
is given in terms of the soft anomalous dimension matrix $\Gamma_S$
\cite{KS,KOS,NKrev}. 
In processes with simple color flow $\Gamma_S$ is a trivial $1\times 1$ 
matrix while in processes with complex 
color flow an appropriate choice of color basis has to be made.
For quark-(anti)quark scattering,  $\Gamma_S$ is a $2\times 2$ 
matrix \cite{KS,BoSt};
for quark-gluon scattering it is a $3\times 3$ matrix \cite{KOS};  
for gluon-gluon scattering it is an $8\times 8$ matrix \cite{KOS}.
For the discussion below we expand $\Gamma_S$ as
$\Gamma_S=(\alpha_s/\pi) \Gamma_S^{(1)}
+(\alpha_s/\pi)^2 \Gamma_S^{(2)}+(\alpha_s/\pi)^3 \Gamma_S^{(3)}
+\cdots$.
The process-dependent soft anomalous dimension matrices have by now been 
presented at one loop for all $2 \rightarrow 2$ 
partonic processes; a compilation of results is given in  \cite{NKrev}. 
They can be explicitly calculated for any process through 
the calculation of eikonal vertex corrections using the techniques and 
results in Refs. \cite{KS,KOS,NKrev}. Some
work has been done on two-loop calculations of these anomalous 
dimensions \cite{NK2l}, and furthermore the universal components of these
anomalous dimensions for quark-antiquark and gluon-gluon
initiated processes have been extracted from the NNLO results 
for Drell-Yan and Higgs production as detailed in 
Ref. \cite{NKuni}. 

The exponentials in the resummed cross section can be expanded 
to any fixed order in $\alpha_s$ and then inverted to momentum space
to provide explicit results for the higher-order corrections.
A fixed-order expansion avoids the problems with infrared singularities 
in the exponents and thus no prescription is needed to deal with these 
in our approach (see discussion in Ref. \cite{NKtop}).

\mysection{NLO master formula for soft-gluon corrections}

We first expand the resummed formula in Eq. (\ref{resHS})
to next-to-leading order and present a master formula 
for the NLO soft-gluon corrections in 
the $\overline{\rm MS}$ scheme and 1PI kinematics:
\beq
{\hat{\sigma}}^{(1)} = \sigma^B \frac{\alpha_s(\mu_R^2)}{\pi}
\left\{c_3\, {\cal D}_1(s_4) + c_2\,  {\cal D}_0(s_4) 
+c_1\,  \delta(s_4)\right\}+\frac{\alpha_s^{d_{\alpha_s}+1}(\mu_R^2)}{\pi} 
\left[A^c \, {\cal D}_0(s_4)+T_1^c \, \delta(s_4)\right] \, ,
\label{NLOmaster}
\eeq
where $\sigma^B$ is the Born term,
\beq
c_3=\sum_i 2 \, C_i -\sum_j C_j\, ,
\label{c3}
\eeq 
with $C_q=C_F$ and $C_g=C_A$,
and $c_2$ is defined by $c_2=c_2^{\mu}+T_2$, 
with
\beq
c_2^{\mu}=-\sum_i C_i \ln\left(\frac{\mu_F^2}{M^2}\right)
\eeq
denoting the terms involving logarithms of the scale, and  
\beq
T_2=- \sum_i \left[C_i
+2 \, C_i \, \ln\left(\frac{-t_i}{M^2}\right)+
C_i \ln\left(\frac{M^2}{s}\right)\right]
-\sum_j \left[B_j^{(1)}+C_j
+C_j \, \ln\left(\frac{M^2}{s}\right)\right] \, 
\label{c2n}
\eeq
denoting the scale-independent terms.
We remind the reader
that the sums over $i$ run over incoming partons and the sums over $j$
run over any massless partons in the final state. 
Note that not all the NLO corrections are proportional to the
Born term; only the leading logarithms and terms involving the scale are.
The function $A^c$ is process-dependent and depends on the color 
structure of the hard-scattering. It is defined by
\beq
A^c={\rm tr} \left(H^{(0)} \Gamma_S^{(1)\,\dagger} S^{(0)}
+H^{(0)} S^{(0)} \Gamma_S^{(1)}\right) \, .
\label{Ac}
\eeq

With regard to the $\delta(x_{th})$ terms, we split them into a term
$c_1$, that is proportional 
to the Born cross section, and a term $T_1^c$ that is not.
$c_1 =c_1^{\mu} +T_1$, with
\beq
c_1^{\mu}=\sum_i \left[C_i\, \ln\left(\frac{-t_i}{M^2}\right) 
-\gamma_i^{(1)}\right]\ln\left(\frac{\mu_F^2}{M^2}\right)
+d_{\alpha_s} \frac{\beta_0}{4} \ln\left(\frac{\mu_R^2}{M^2}\right) 
\label{c1mu}
\eeq
denoting the terms involving logarithms of the scale.
$T_1$ and $T_1^c$ do not involve the factorization and 
renormalization scales.
Note that $T_1$ and $T_1^c$ are virtual terms and cannot be derived from 
the resummation formalism, but they can be read off
by matching to a full NLO calculation for any specified process.

In PIM kinematics we simply replace $s_4$ by $1-z$, set $s=M^2$, and 
delete all $\ln(-t_i/M^2)$ terms from the above expressions. 
The same should be done for the NNLO and NNNLO results that follow.

As shown in Ref. \cite{NKuni} the NLO master formula passes a number of tests. 
Its predictions agree with NLO soft-gluon results for all 
processes where those results are already available. Also, the renormalization 
and factorization scale dependence in the physical cross section 
(after convoluting the partonic cross section with the parton distributions)
cancels out explicitly, i.e. $d\sigma/d\mu_F=0$ and
$d\sigma/d\mu_R=0$ at NLO.

\mysection{NNLO master formula for soft-gluon corrections}

At next-to-next-to-leading order, the expansion of Eq. (\ref{resHS}), with 
matching to the NLO soft-plus-virtual result, Eq. (\ref{NLOmaster}), 
gives the NNLO soft corrections in the $\overline{\rm MS}$ scheme and
1PI kinematics:
\beqa
{\hat{\sigma}}^{(2)}&=& 
\sigma^B \frac{\alpha_s^2(\mu_R^2)}{\pi^2} \;
\frac{1}{2} \, c_3^2 \; {\cal D}_3(s_4)
\nonumber \\ && \hspace{-15mm}
{}+\sigma^B \frac{\alpha_s^2(\mu_R^2)}{\pi^2} \;
\left\{\frac{3}{2} \, c_3 \, c_2 - \frac{\beta_0}{4} \, c_3
+\sum_j C_j \, \frac{\beta_0}{8}\right\} \; {\cal D}_2(s_4)
+\frac{\alpha_s^{d_{\alpha_s}+2}(\mu_R^2)}{\pi^2} \; 
\frac{3}{2} \, c_3 \, A^c\; {\cal D}_2(s_4)
\nonumber \\ && \hspace{-15mm}
{}+\sigma^B \frac{\alpha_s^2(\mu_R^2)}{\pi^2} \; C_{D_1}^{(2)} \;  
{\cal D}_1(s_4) 
+\frac{\alpha_s^{d_{\alpha_s}+2}(\mu_R^2)}{\pi^2} \; 
\left\{\left(2\, c_2-\frac{\beta_0}{2}\right)\, A^c+c_3 \, T_1^c
+F^c\right\} \; {\cal D}_1(s_4)
\nonumber \\ && \hspace{-15mm} 
{}+\sigma^B \frac{\alpha_s^2(\mu_R^2)}{\pi^2} \; C_{D_0}^{(2)} \; 
{\cal D}_0(s_4)  
\nonumber \\ &&  \hspace{-15mm}
{}+\frac{\alpha_s^{d_{\alpha_s}+2}(\mu_R^2)}{\pi^2} \; 
\left\{\left[c_1-\zeta_2 \, c_3+\frac{\beta_0}{4} \, 
\ln\left(\frac{\mu_R^2}{M^2}\right)
+\frac{\beta_0}{4} \, 
\ln\left(\frac{M^2}{s}\right)\right]\, A^c
+\left(c_2-\frac{\beta_0}{2}\right) \, T_1^c 
\right.
\nonumber \\ && \hspace{20mm} \left.
{}+F^c \, \ln\left(\frac{M^2}{s}\right)+G^c\right\} \;  
{\cal D}_0(s_4)
\nonumber \\ &&  \hspace{-15mm}
{}+\sigma^B \frac{\alpha_s^2(\mu_R^2)}{\pi^2} \; R^{(2)} \; \delta(s_4) 
+\frac{\alpha_s^{d_{\alpha_s}+2}(\mu_R^2)}{\pi^2} \; R_c^{(2)} \; 
\delta(s_4) \, .
\label{NNLOmaster}
\eeqa
We have used the definitions
\beq
C_{D_1}^{(2)}=c_3 \, c_1 +c_2^2
-\zeta_2 \, c_3^2 -\frac{\beta_0}{2} \, T_2 
+\frac{\beta_0}{4} \, c_3 \, \ln\left(\frac{\mu_R^2}{M^2}\right)
+c_3\, \frac{K}{2} 
-\sum_j\frac{\beta_0}{4} \, B_j^{(1)} \, , 
\eeq
\beqa
C_{D_0}^{(2)}&=&c_2 \, c_1 -\zeta_2 \, c_3 \, c_2+\zeta_3 \, c_3^2 
-\frac{\beta_0}{2} \, T_1
+\frac{\beta_0}{4}\, c_2 \, \ln\left(\frac{\mu_R^2}{M^2}\right) 
+d_{\alpha_s}\frac{\beta_0^2}{8}\ln\left(\frac{M^2}{s}\right)
-\sum_i {\nu}_i^{(2)}  
\nonumber \\ && 
{}+\sum_i C_i \, \frac{\beta_0}{8} \left[ 
\ln^2\left(\frac{\mu_F^2}{M^2}\right)
-\ln^2\left(\frac{M^2}{s}\right)
-2\, \ln\left(\frac{M^2}{s}\right) \right]
-\frac{\beta_0}{2}\sum_i \gamma_i^{(1)} \ln\left(\frac{M^2}{s}\right)
\nonumber \\ &&
{}-\sum_i C_i\, \frac{K}{2} \left[\ln\left(\frac{\mu_F^2}{M^2}\right)
+2\, \ln\left(\frac{-t_i}{M^2}\right)
+\ln\left(\frac{M^2}{s}\right)\right]
-\sum_j \left(B_j^{(2)}+\nu_j^{(2)}\right)
\nonumber \\ && \hspace{-10mm} \quad \quad
+\sum_j C_j \, \left[-\frac{\beta_0}{8} \, 
\ln^2\left(\frac{M^2}{s}\right)
-\frac{\beta_0}{4} \, \ln\left(\frac{M^2}{s}\right)
-\frac{K}{2} \, \ln\left(\frac{M^2}{s}\right)\right]
-\sum_j \frac{\beta_0}{2} \, 
B_j^{(1)} \, \ln\left(\frac{M^2}{s}\right) \, ,
\nonumber \\
\eeqa
\beq
F^c={\rm tr} \left[H^{(0)} \left(\Gamma_S^{(1)\,\dagger}\right)^2 S^{(0)}
+H^{(0)} S^{(0)} \left(\Gamma_S^{(1)}\right)^2
+2 H^{(0)} \Gamma_S^{(1)\,\dagger} S^{(0)} \Gamma_S^{(1)} \right] \, ,
\label{Fterm}
\eeq
\beqa
G^c&=&{\rm tr} \left[H^{(1)} \Gamma_S^{(1)\,\dagger} S^{(0)}
+H^{(1)} S^{(0)} \Gamma_S^{(1)} + H^{(0)} \Gamma_S^{(1)\,\dagger} S^{(1)}
+H^{(0)} S^{(1)} \Gamma_S^{(1)} \right.
\nonumber \\ && \quad \quad \left.
{}+H^{(0)} \Gamma_S^{(2)\,\dagger} S^{(0)}
+H^{(0)} S^{(0)} \Gamma_S^{(2)} \right] \, .
\eeqa
In PIM kinematics simply replace $s_4$ by $1-z$, set $s=M^2$, and 
delete all $\ln(-t_i/M^2)$ terms. 

The quantities $d_{\alpha_s}$, $\beta_0$, and $K$
have all been defined in Section 2, and $\zeta_2=\pi^2/6$, 
$\zeta_3=1.2020569\cdots$.
Also $c_3$, $c_2$, $c_1$, $T_1^c$, and $A^c$ have been defined in Section 3.
The virtual terms $R^{(2)}$ and $R_c^{(2)}$ cannot be derived from 
resummation. A separate calculation is needed for each process to derive 
those. However, all the scale-dependent terms in $R^{(2)}$  ($R_c^{(2)}$  
is scale-independent) can be derived and are given explicitly in 
Eq. (12) of Ref. \cite{NKmpla}. 

As shown in Ref. \cite{NKuni}, the NNLO master formula passes many rigorous
tests. It reproduces the NNLO soft-gluon results 
for all processes where these results are known.
Also at NNLO the renormalization and
factorization scale dependence in the physical cross section cancels out.

The master formula can in principle provide all the soft corrections 
at NNLO for any process. In practice, 
the accuracy which we can attain depends on whether the one-loop 
$\Gamma_S^{(1)}$ is known (in which case we can attain NLL accuracy;
$\Gamma_S^{(1)}$ is known for all 2 $\rightarrow$ 2 processes); whether
furthermore the NLO virtual terms are known (NNLL accuracy); and 
whether the two-loop $\Gamma_S^{(2)}$ is known (NNNLL accuracy).
Most current results are known to NLL or NNLL accuracy.
Note that $\Gamma_S^{(2)}$ is only known for the simplest cases of
Drell-Yan and Higgs production where the color structure is trivial
\cite{NKuni}.
However, it was shown in \cite{NKRVtop} that the contributions of
$\Gamma_S^{(2)}$ can be small so that effectively NNNLL calculations 
can be made in some cases even when $\Gamma_S^{(2)}$ is not fully known.

\mysection{NNNLO master formula for soft-gluon corrections}

At next-to-next-to-next-to-leading order, the expansion of 
Eq. (\ref{resHS}), with 
matching to the NLO and NNLO soft-plus-virtual results, gives the NNNLO
soft-gluon corrections in the $\overline{\rm MS}$ scheme and
1PI kinematics
\beqa
{\hat{\sigma}}^{(3)}&=& 
\sigma^B \frac{\alpha_s^3(\mu_R^2)}{\pi^3} \;  
\frac{1}{8} \, c_3^3 \; {\cal D}_5(s_4)
\nonumber \\ && \hspace{-10mm}
{}+\sigma^B \frac{\alpha_s^3(\mu_R^2)}{\pi^3} \; 
\left\{\frac{5}{8} \, c_3^2 \, c_2 -\frac{5}{2} \, c_3 \, X_3\right\} \;  
{\cal D}_4(s_4)
+\frac{\alpha_s^{d_{\alpha_s}+3}(\mu_R^2)}{\pi^3} \;
\frac{5}{8} \, c_3^2 \, A^c \; {\cal D}_4(s_4)
\nonumber \\ && \hspace{-10mm}
{}+\sigma^B \frac{\alpha_s^3(\mu_R^2)}{\pi^3} \;  
\left\{c_3 \, c_2^2 +\frac{1}{2} \, c_3^2 \, c_1
-\zeta_2 \, c_3^3 +(\beta_0-4\, c_2) \, X_3 +2 \, c_3 \, X_2
-\sum_j C_j \, \frac{\beta_0^2}{48}\right\} \; {\cal D}_3(s_4)
\nonumber \\ && \hspace{-10mm}
{}+\frac{\alpha_s^{d_{\alpha_s}+3}(\mu_R^2)}{\pi^3} \;
\left\{\frac{1}{2}\, c_3^2\, T_1^c+\left[2\, c_3 \, c_2
-\frac{\beta_0}{2} \, c_3 -4\,  X_3 \right] \,A^c +c_3 \, F^c\right\} \;
{\cal D}_3(s_4)
\nonumber \\ && \hspace{-10mm} 
{}+\sigma^B \frac{\alpha_s^3(\mu_R^2)}{\pi^3} \;
\left\{\frac{3}{2}\, c_3 \,c_2 \, c_1 +\frac{1}{2} \, c_2^3
-3\, \zeta_2 \, c_3^2 \,c_2 +\frac{5}{2} \, \zeta_3 \, c_3^3
+\left(-3 \, c_1+\frac{27}{2} \, \zeta_2 \, c_3\right) \, 
X_3 \right.
\nonumber \\ && \hspace{-5mm}
{}+(3\, c_2-\beta_0) \, X_2 -\frac{3}{2} \, c_3 \, X_1
-\sum_i C_i \, \frac{\beta_1}{8}+\sum_j C_j \, \frac{\beta_0}{16} \, 
\left[\beta_0 \, \ln\left(\frac{\mu_R^2}{M^2}
\right)+2\, K\right]
\nonumber \\ && \hspace{-5mm} \left.
{}+\sum_j \frac{\beta_0^2}{16} \, {B'}_j^{(1)}+\sum_j\frac{3}{32}\, C_j \,
\beta_1 \right\} \; {\cal D}_2(s_4)  
\nonumber \\ && \hspace{-10mm}
{}+\frac{\alpha_s^{d_{\alpha_s}+3}(\mu_R^2)}{\pi^3} \;
\left\{\left(\frac{3}{2}\, c_3 \, c_2-3\, X_3\right) \, T_1^c 
+\frac{3}{2}\, \left[c_2+c_3 \, \ln\left(\frac{M^2}{s}\right)\right]\, F^c
\right.
\nonumber \\ && \hspace{-5mm}
{}+\left[\frac{3}{2}\, c_2^2+\frac{3}{2}\, c_3 \, c_1
-3 \, \zeta_2 \, c_3^2 +3 \, X_2
+\frac{\beta_0^2}{4}-\frac{3}{4}\, \beta_0 \, \left(c_2-\frac{c_3}{2} \, 
\ln\left(\frac{\mu_R^2}{M^2}\right)
+\frac{c_3}{2}\, \ln\left(\frac{M^2}{s}\right)\right)\right] \, A^c
\nonumber \\ && \hspace{5mm} \left.
{}+\frac{3}{2} \, c_3 \, G^c+\frac{1}{2} \, K_3^c\right\} \;
{\cal D}_2(s_4)
\nonumber \\ &&  \hspace{-10mm}
{}+\sigma^B \frac{\alpha_s^3(\mu_R^2)}{\pi^3} \;
\left\{\frac{1}{2}\, c_3 \, c_1^2 +c_2^2 \, c_1
-\zeta_2 \, c_3^2 \, c_1-\frac{5}{2}\, \zeta_2 \, c_3 \, c_2^2
+5 \, \zeta_3 \, c_3^2 \, c_2
+\frac{5}{4}\, \zeta_2^2 \, c_3^3 -\frac{15}{4}\, \zeta_4 \, c_3^3 \right.
\nonumber \\ && \hspace{-5mm}
{}-\frac{\beta_0^2}{4}\, \zeta_2 \, c_3
+(-20\, \zeta_3 \, c_3+12 \, \zeta_2 \, c_2) \, X_3 
+(2\, c_1-5\, \zeta_2 \, c_3) \, X_2 
+(\beta_0- 2 \, c_2) \, X_1 + c_3\, X_0
\nonumber \\ && \hspace{-5mm} \left.
{}+\sum_i Y_i^{(1)} +\sum_j Y_j^{(1)}
\right\} \; {\cal D}_1(s_4)
\nonumber \\ && \hspace{-10mm}
{}+\frac{\alpha_s^{d_{\alpha_s}+3}(\mu_R^2)}{\pi^3} \;
\left\{C^{(2)}_{D_1} \, \left[T_1^c-A^c \,\ln\left(\frac{M^2}{s}\right)\right] 
+c_3 \, {\rm tr} \left[H^{(2)}S^{(0)}+H^{(0)}S^{(2)}+H^{(1)}S^{(1)}\right]
\right.
\nonumber \\ && \hspace{-5mm}
{}+\left[2\, c_2 \, c_1 -5 \, \zeta_2 \, c_3 \, c_2 +5 \, \zeta_3 \, c_3^2
+12 \, \zeta_2 \, X_3 -2 \, X_1+\ln\left(\frac{M^2}{s}\right) \,
\left(c_2^2+c_3 \, c_1-\zeta_2 \, c_3^2+2 \, X_2\right) \right. 
\nonumber \\ &&
{}-\frac{\beta_0}{2}\left(c_1-\frac{5}{2}\, \zeta_2 \, c_3
-c_2\,\ln\left(\frac{\mu_R^2}{M^2}\right)
-\frac{c_3}{2}\, \ln\left(\frac{M^2}{s}\right) \, 
\ln\left(\frac{\mu_R^2}{M^2}\right) 
+c_2 \, \ln\left(\frac{M^2}{s}\right)\right)
\nonumber \\ && \left. 
{}+\frac{\beta_0^2}{4}\, \left(\ln\left(\frac{M^2}{s}\right)
- \ln\left(\frac{\mu_R^2}{M^2}\right)\right)-\frac{\beta_1}{8}
\right] \, A^c
\nonumber \\ && \hspace{-5mm}
{}+\left[-\frac{5}{2} \, \zeta_2 \, c_3 +c_1+2\, c_2 \,  
\ln\left(\frac{M^2}{s}\right)+\frac{c_3}{2}\, 
\ln^2\left(\frac{M^2}{s}\right)-\frac{\beta_0}{2}\right] \, F^c
\nonumber \\ && \hspace{-5mm} \left.
{}+\left[c_3 \, \ln\left(\frac{M^2}{s}\right)+2\, c_2\right] \, G^c
-\beta_0 \, \left(G^c-M^c\right)
+K_2^c+K_3^c \, \ln\left(\frac{M^2}{s}\right) \right\} \; 
{\cal D}_1(s_4)
\nonumber \\ && \hspace{-10mm}
{}+\sigma^B \frac{\alpha_s^3(\mu_R^2)}{\pi^3} \;  
\left\{\frac{1}{2}\, c_2 \, c_1^2+3 \, \zeta_5 \, c_3^3
-\frac{15}{4} \, \zeta_4 \, c_3^2 \, c_2
-2 \, \zeta_2 \, \zeta_3 \, c_3^3 +\zeta_3 \, c_3^2 \, c_1
+ 2 \, \zeta_3 \, c_3 \, c_2^2
+\frac{5}{4} \, \zeta_2^2 \, c_3^2 \, c_2 
\right.
\nonumber \\ && \hspace{-5mm}
{}-\zeta_2 \, c_3 \, c_2 \, c_1-\frac{1}{2} \, \zeta_2 \, c_2^3
+(15 \, \zeta_4 \, c_3-8 \, \zeta_3 \, c_2-6 \, \zeta_2^2 \, c_3 
+3 \, \zeta_2 \, c_1) \, X_3
+(4 \, \zeta_3 \, c_3 -3 \, \zeta_2 \, c_2) \, X_2
\nonumber \\ && \hspace{-5mm} \left.
{}+(\zeta_2 \, c_3-c_1)\, X_1
+c_2 \, X_0-\frac{\beta_0^2}{4} \, T_1 \, \ln\left(\frac{\mu_R^2}{M^2}\right)
+\frac{\beta_0^2}{16} \, T_2 \, \ln^2\left(\frac{\mu_R^2}{M^2}\right)
-\frac{\beta_0^2}{4}\, T_1\, \ln\left(\frac{M^2}{s}\right) \right.
\nonumber \\ &&  \hspace{-5mm} \left.
{}-\frac{\beta_1}{8} \, T_1 +\sum_i Y_i^{(0)} +\sum_j Y_j^{(0)}\right\} \; 
{\cal D}_0(s_4) 
\nonumber \\ && \hspace{-10mm}
{}+\frac{\alpha_s^{d_{\alpha_s}+3}(\mu_R^2)}{\pi^3} \;
\left\{C^{(2)}_{D_0} \, \left[T_1^c-A^c \,\ln\left(\frac{M^2}{s}\right)\right]
{}+\left[c_2 \, \ln\left(\frac{M^2}{s}\right)-\zeta_2 \, c_3+c_1\right]\, G^c
\right.
\nonumber \\ && \hspace{-5mm}
{}+\left[\frac{1}{2}\, \left(c_1+\frac{\zeta_2}{2}c_3\right)^2
+\frac{9}{8}\, \zeta_2^2 \, c_3^2-\frac{15}{4} \zeta_4 \, c_3^2
+4 \, \zeta_3 \, c_3 \, c_2-\frac{3}{2}\, \zeta_2 \, c_3 \, c_1
-\frac{3}{2} \, \zeta_2 \, c_2^2 -8 \, \zeta_3 \, X_3 \right.
\nonumber \\ && \left.
{}-3\, \zeta_2 \, X_2 + X_0
+\left(c_2\, c_1-\zeta_2 \, c_3 \, c_2 
+\zeta_3 \, c_3^2+3\, \zeta_2 \, X_3 - X_1 \right) \, 
\ln\left(\frac{M^2}{s}\right)\right]\, A^c
\nonumber \\ && \hspace{-5mm}
{}-\frac{\beta_0}{2}\, \left[-\frac{c_2}{2} \,
\ln\left(\frac{M^2}{s}\right)\,\ln\left(\frac{\mu_R^2}{M^2}\right)
+2\, \zeta_3\, c_3-\frac{3}{2} \, \zeta_2 \, c_2 \right.
\nonumber \\ && \hspace{10mm} \left.
{}+\frac{(c_1-\zeta_2\, c_3)}{2}\, 
\left(-\ln\left(\frac{\mu_R^2}{M^2}\right)
+\ln\left(\frac{M^2}{s}\right)\right)\right]\, A^c
\nonumber \\ && \hspace{-5mm}
{}+\left[\frac{c_2}{2}\, \ln^2\left(\frac{M^2}{s}\right)
+\left(c_1-\zeta_2\, c_3-\frac{\beta_0}{4}\right) \, 
\ln\left(\frac{M^2}{s}\right)
+2\, \zeta_3 \, c_3-\frac{3}{2}\, \zeta_2 \, c_2
+\frac{\beta_0}{4}\ln\left(\frac{\mu_R^2}{M^2}\right)\right] \, F^c
\nonumber \\ && \hspace{-5mm}
{}+\left[\frac{\beta_0^2}{16}\, \left(
-2\, \ln\left(\frac{M^2}{s}\right)\, \ln\left(\frac{\mu_R^2}{M^2}\right)
+\ln^2\left(\frac{\mu_R^2}{M^2}\right)
+\ln^2\left(\frac{M^2}{s}\right)
-4\, \zeta_2\right)\right.
\nonumber \\ && \left. 
{}-\frac{\beta_1}{16}\, \left(\ln\left(\frac{M^2}{s}\right)
-\ln\left(\frac{\mu_R^2}{M^2}\right)\right)\right] \, M^c
+c_2 \, {\rm tr} \left[H^{(2)}S^{(0)}+H^{(0)}S^{(2)}+H^{(1)}S^{(1)}\right]
\nonumber \\ && \hspace{-5mm} \left.
{}+K_1^c+K_2^c \, \ln\left(\frac{M^2}{s}\right)
+\frac{1}{2}\left[-\zeta_2+ \ln^2\left(\frac{M^2}{s}\right)\right]\, K_3^c
\right\} \; {\cal D}_0(s_4) \, .
\label{NNNLOmaster}
\eeqa
The quantities $d_{\alpha_s}$, $\beta_0$, and $K$
have been defined in Section 2, while
$c_3$, $c_2$, $c_1$, $T_1^c$, and $A^c$ have been defined in Section 3.
Also $C^{(2)}_{D_1}$, $C^{(2)}_{D_0}$, $F^c$, $G^c$, $\zeta_2$, and $\zeta_3$
have been defined in Section 4, and $\zeta_4=\pi^4/90$.
The expressions for $\beta_1$, $X_3$, $X_2$, $X_1$, $X_0$, $Y_i^{(1)}$, 
$Y_j^{(1)}$, $Y_i^{(0)}$, $Y_j^{(0)}$, $M^c$, $K_1^c$, $K_2^c$, and
$K_3^c$ are given in the Appendix.

Note again that in PIM kinematics we simply replace $s_4$ by $1-z$,
set $s=M^2$, and drop the terms with  
$\ln(-t_i/M^2)$ in the above formula and in all the expressions
given in the Appendix. 

This NNNLO master equation 
gives the structure of the NNNLO soft corrections and can provide 
the full soft corrections explicitly if all the two-loop and three-loop 
quantities are known. Therefore for processes 
with non-trivial color structure we are currently limited to NLL or 
NNLL accuracy, as the applications to charged Higgs  
and top quark production in the next two sections illustrate.
The structure of the corrections as presented here can be useful for 
checking future calculations if and when such three-loop quantities 
become available. Also note that scale logarithms and $\zeta_i$ 
constants will be kept as appropriate at subleading logs as explained 
in the next two sections. 

When the color structure of the hard scattering is simple, i.e. 
when $H$, $S$, and $\Gamma_S$ are simply $1\times 1$ matrices,
then the above expressions can be simplified. We can then easily absorb
$A^c$ into $c_2$ in Eq. (\ref{NLOmaster}), and $T_1^c$ into $c_1$,
by redefining $c_2$ and $c_1$.
Then all the terms are proportional to $\sigma^B$ in Eqs. (\ref{NLOmaster}),
(\ref{NNLOmaster}), and (\ref{NNNLOmaster}). We will see this
explicitly in the two applications in the next two sections.  

\mysection{Charged Higgs production via $bg \rightarrow tH^-$}

Charged Higgs production is a process of great interest at the 
LHC. The charged Higgs boson, if discovered, would be an unmistakable 
sign of new physics beyond the Standard Model \cite{Higgs}. 
A promising channel of discovery is associated production with a top quark 
via bottom-gluon fusion for which SUSY and QCD radiative corrections have been
calculated \cite{BGGS,Zhu,Plehn,BHJP}. 
NLO and NNLO soft-gluon corrections to this process were 
recently studied in \cite{NKchiggs} where the corrections were found 
to be large, especially for a very massive charged Higgs.

We now apply our  NNNLO master formula to charged Higgs production 
in the Minimal Supersymmetric Standard Model (MSSM)
via bottom gluon fusion, a process with simple color flow, at NLL accuracy.
We study the process $b(p_b)+g(p_g) \rightarrow t(p_t)+H^-(p_{H^-})$ 
in 1PI kinematics and define the kinematical invariants
$s=(p_b+p_g)^2$, $t=(p_b-p_t)^2$, $u=(p_g-p_t)^2$, and 
$s_4=s+t+u-m_t^2-m_{H^-}^2$, where $m_t$ is the top quark mass and
$m_{H^-}$ is the charged Higgs mass (we take the bottom quark $m_b=0$
in the kinematics \cite{NKchiggs}) .

The NLO coefficients of Section 3 here take the values
$c_3=2(C_F+C_A)$ and $c_2=c_2^{\mu}+T_2$ with
$c_2^{\mu}=-(C_F+C_A)
\ln(\mu_F^2/m_{H^-}^2)$ and
\beqa
T_2&=&2 {\rm Re} \Gamma_S^{(1)}
-C_F-C_A-2C_F\ln\left(\frac{-u+m^2_{H^-}}{m^2_{H^-}}\right)
-2C_A\ln\left(\frac{-t+m^2_{H^-}}{m^2_{H^-}}\right)
\nonumber \\ &&
{}-(C_F+C_A)\ln\left(\frac{m_{H^-}^2}{s}\right) \, , 
\eeqa
where ${\rm Re} \Gamma_S^{(1)}$ denotes the real part of the one-loop 
soft anomalous dimension \cite{NKchiggs},
\beq
\Gamma_S^{(1)}=C_F \ln\left(\frac{-t+m_t^2}{m_t\sqrt{s}}\right)
+\frac{C_A}{2} \ln\left(\frac{-u+m_t^2}{-t+m_t^2}\right)
+\frac{C_A}{2} (1-\pi i) \, .
\eeq
Note that, due to the simple 
color structure of this process, $\Gamma_S$ is simply a $1\times 1$ matrix.
Here, as described in the last paragraph of the previous section,
we have aborbed the term $2 {\rm Re} \Gamma_S^{(1)}$, which arises
from $A^c$, into $T_2$.
Also,
\beqa
c_1^{\mu}&=&\left[C_F \ln\left(\frac{-u+m^2_{H^-}}
{m^2_{H^-}}\right)
+C_A \ln\left(\frac{-t+m^2_{H^-}}{m^2_{H^-}}\right)
-\frac{3}{4}C_F-\frac{\beta_0}{4}\right]
\ln\left(\frac{\mu_F^2}{m^2_{H^-}}\right)
\nonumber \\ &&
{}+\frac{\beta_0}{4} \ln\left(\frac{\mu_R^2}{m^2_{H^-}}\right) \, .
\eeqa

The various terms from the Appendix used in the NNNLO corrections 
here take the values  
$X_3=\beta_0 \, c_3/12$,  
$X_2^{\mu}=(\beta_0/8)c_3\ln(\mu_R^2/m_{H^-}^2)$,
\beq
X_1^{\mu^2}=-\frac{\beta_0}{4}c_2^{\mu}
\ln\left(\frac{\mu_R^2}{m_{H^-}^2}\right)
-(C_F+C_A)\frac{\beta_0}{8}\ln^2\left(\frac{\mu_F^2}{m_{H^-}^2}\right)\, ,
\eeq
\beq
X_1^{\zeta}=\frac{\beta_0}{4}\, \zeta_2 \, c_3 \, ,
\eeq
where $X_2^{\mu}$ denotes the scale logarithm terms in $X_2$,
$X_1^{\mu^2}$ denotes terms involving squares of the scale logarithms in $X_1$,
and $X_1^{\zeta}$ denotes the $\zeta_i$ terms in $X_1$.

\begin{figure}
\begin{center}
\includegraphics[width=12.5cm]{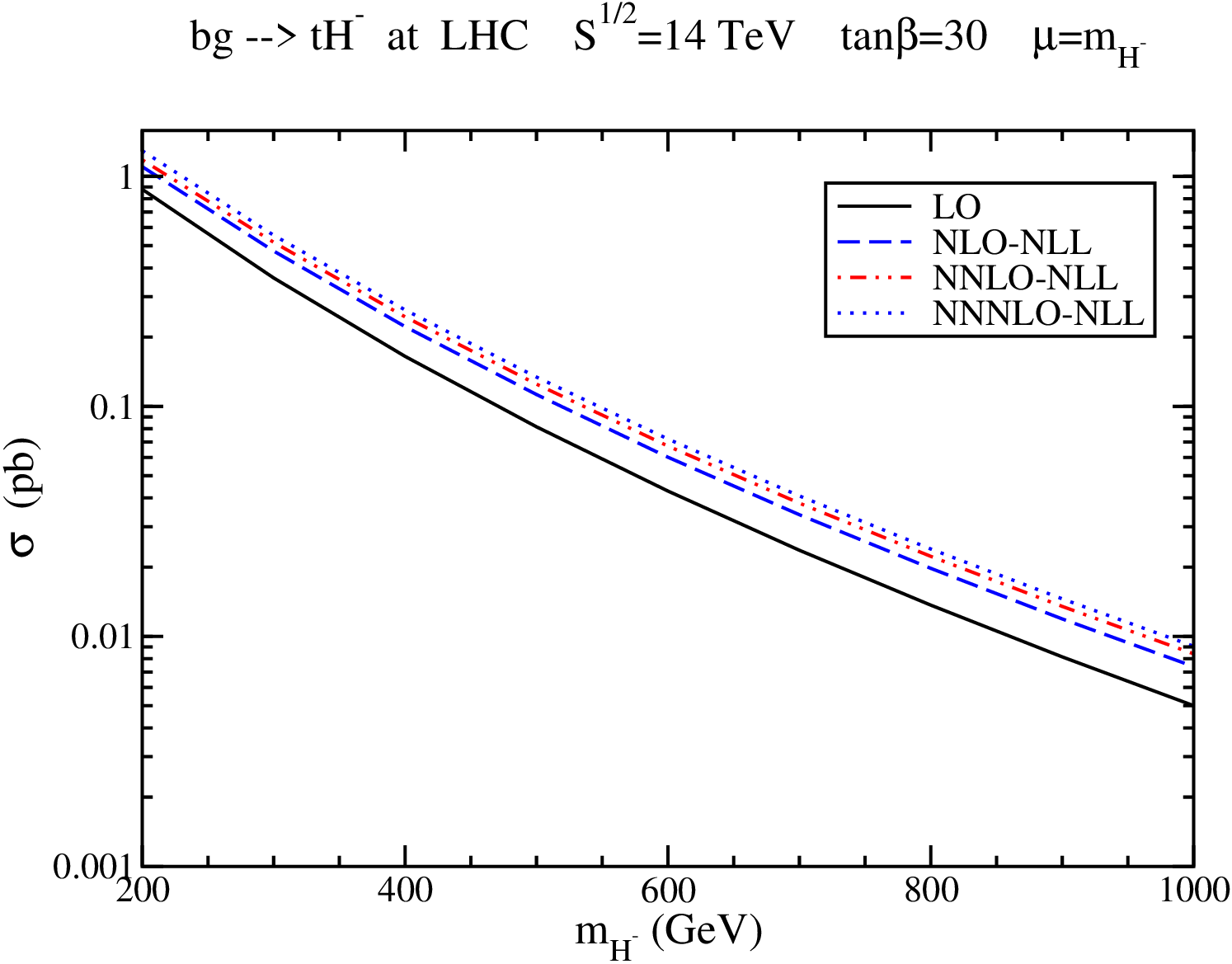}
\caption{The total cross section for charged Higgs production at the LHC.}
\label{higgs3amhplot}
\end{center}
\end{figure}

The NNNLO-NLL corrections are then given by 
\beqa
{\hat{\sigma}}^{(3)}&=& 
\sigma^B \frac{\alpha_s^3(\mu_R^2)}{\pi^3} \;  
\left\{ \frac{1}{8} \, c_3^3 \; {\cal D}_5(s_4)
+ \left[\frac{5}{8} \, c_3^2 \, c_2 -\frac{5}{2} \, c_3 \, X_3\right] \;  
{\cal D}_4(s_4) \right.
\nonumber \\ && \hspace{-10mm}
{}+\left[c_3 \, (c_2^{\mu})^2+2 \, c_3 \, T_2 \, c_2^{\mu} 
+\frac{1}{2} \, c_3^2 \, c_1^{\mu}
-\zeta_2 \, c_3^3 -4\, c_2^{\mu} \, X_3 +2 \, c_3 \, X_2^{\mu}
\right] \; {\cal D}_3(s_4)
\nonumber \\ && \hspace{-10mm} 
{}+\left[\frac{3}{2}\, c_3 \,c_2^{\mu} \, c_1^{\mu} +\frac{1}{2} \, 
(c_2^{\mu})^3+\frac{3}{2} T_2 (c_2^{\mu})^2-3\, \zeta_2 \, c_3^2 \,c_2 
+\frac{5}{2} \, \zeta_3 \, c_3^3 \right.
\nonumber \\ && \left. 
{}+\frac{27}{2}\, \zeta_2 \, c_3 \, X_3+3\, c_2^{\mu} \, X_2^{\mu} 
-\frac{3}{2} \, c_3 \, (X_1^{\mu^2}+X_1^{\zeta})\right] \; {\cal D}_2(s_4)  
\nonumber \\ && \hspace{-10mm} 
{}+\left[(c_2^{\mu})^2\, c_1^{\mu}-\zeta_2 \, c_3^2\, c_1^{\mu}
-\frac{5}{2} \, \zeta_2 \, c_3 \, \left((c_2^{\mu})^2+2\, T_2 \, 
c_2^{\mu}\right)+5\, \zeta_3 \, c_3^2 \, c_2^{\mu}\right.
\nonumber \\ && \left. \left.
{}+12\, \zeta_2 \, c_2^{\mu} \, X_3-5\, \zeta_2\, c_3\, X_2^{\mu}
-2\, c_2^{\mu}\, \left(X_1^{\mu^2}+X_1^{\zeta}\right)
\right] \; {\cal D}_1(s_4) \right\} \, .
\label{NNNLOchiggs}
\eeqa
Because we absorbed $A^c$ into $c_2$ the corrections take 
a simple form, simply multiplying the Born term $\sigma^B$.
Note that consistent with a NLL calculation we include all
${\cal D}_5$ (LL) and ${\cal D}_4$ (NLL) terms. In addition, we calculate 
all scale logarithms at NLL accuracy. 
This means that for coefficients of 
$\ln^i(\mu^2/m^2)$ we include the most singular plus distribution
and the next-most-singular one \cite{KLMV}. Thus, we also include
all scale logarithms in the ${\cal D}_3$ terms, the cubed and 
squared scale logarithms in the ${\cal D}_2$ terms, and the cubed 
scale logarithms in the ${\cal D}_1$ terms.  
With respect to the subleading $\zeta_i$ terms that arise from inversion from 
moment to momentum space, we include only those that we can calculate 
exactly (for a discussion of the numerical effects of such terms 
see Ref. \cite{NKtop}). Thus we include all $\zeta_i$ terms
in the ${\cal D}_3$ and ${\cal D}_2$ terms, and all $\zeta_i$ terms multiplying
scale logarithms in the ${\cal D}_1$ term.

In Fig. \ref{higgs3amhplot} we plot the cross section versus
charged Higgs mass for $pp$ collisions at the LHC with $\sqrt{S}=14$ TeV 
using the MRST2002 approximate
NNLO parton distribution functions \cite{MRST2002}
with the respective  three-loop evaluation of $\alpha_s$.
We set the factorization scale equal to the renormalization scale
and denote this common scale by $\mu$.
We show results for the LO, NLO-NLL, NNLO-NLL, and NNNLO-NLL
cross sections, all with a choice of scale $\mu=m_{H^-}$.
In the calculation we choose a value $\tan \beta=30$; here 
$\tan \beta$ is the ratio of the vacuum expectation values of the
two Higgs doublets in the MSSM.
It is straightforward to calculate results for any other value of
$\tan \beta$, since the only dependence on $\beta$ 
is in a factor $m_b^2 \tan^2 \beta+m_t^2 \cot^2 \beta$ appearing
in the Born term.
The cross sections span over two orders of magnitude in the mass
range shown, 200 GeV $\le m_{H^-} \le$ 1000 GeV.
The NLO, NNLO, and NNNLO threshold corrections are positive and provide 
a significant enhancement to the lowest-order result.
We note that the cross sections for the related process
${\bar b} g \rightarrow {\bar t} H^+$ in the MSSM are exactly the same.

\begin{figure}
\begin{center}
\includegraphics[width=12.5cm]{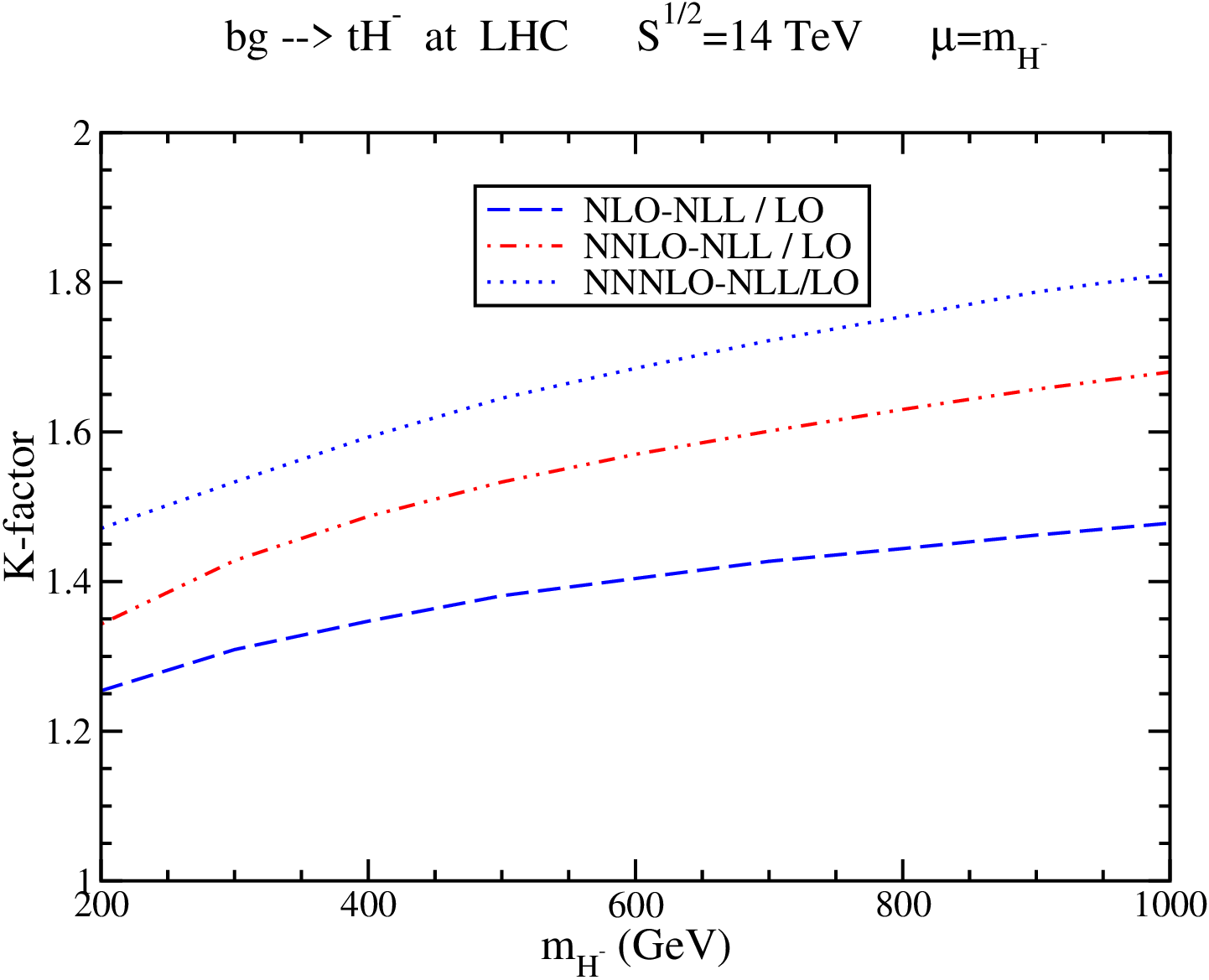}
\caption{The $K$-factors for charged Higgs production at the LHC.}
\label{Khiggs3amhplot}
\end{center}
\end{figure}

In Fig. \ref{Khiggs3amhplot} we plot the $K$-factors, i.e. ratios of cross
sections at higher orders to the LO result, to better show the 
relative size of the corrections. The NLO-NLL / LO curve shows that the
NLO soft-gluon corrections enhance the LO cross section by approximately
25\% to 50\% depending on the mass of the charged Higgs. As expected
the corrections increase for higher charged Higgs masses as 
we get closer to threshold.
The NNLO-NLL / LO curve shows that if we include the NNLO threshold
corrections we get an enhancement over the LO result of approximately
35\% to 70\% in the range of masses shown. Again the enhancement increases
with charged Higgs mass, as expected.
Finally, the NNNLO-NLL / LO curve shows  the further enhancement
that the NNNLO soft-gluon corrections provide, approximately 45\% to 80\% 
over LO. We note that the NNNLO soft-gluon corrections are quite 
significant in this case. This is not necessarily typical for
other processes, but happens here 
because of the very massive final state. Another process where NNNLO
soft-gluon corrections are known to be big is inclusive hadron production 
at high transverse momentum \cite{dFV}.

\mysection{Top quark production via $q{\bar q} \rightarrow t{\bar t}$}

The study of the top quark is important in understanding the electroweak
sector and searching for new physics. The top quark 
is now being actively studied at Run II at the Tevatron \cite{CDF,D0,Wagner}.
Theoretically, the production cross section has been studied to
NNLO and NNNLL in both 1PI and PIM kinematics \cite{NKRVtop}. The corrections 
are moderate and they substantially decrease the scale dependence 
of the cross section. Transverse momentum distributions are also
known to NNLO-NNNLL \cite{NKRVtop}, while rapidity distributions have
been presented in Ref. \cite{KSpty}.

Here we apply our  NNNLO master formula to 
top quark pair production via quark-antiquark annihilation, 
which is the dominant partonic subprocess at the Tevatron.
We study the channel
$q(p_a)+{\bar q}(p_b) \rightarrow t(p_1)+{\bar t}(p_2)$ 
in 1PI kinematics, and define the kinematical invariants 
$s=(p_a+p_b)^2$, $t_1=(p_b-p_1)^2-m_t^2$, $u_1=(p_a-p_1)^2-m_t^2$,
and $s_4=s+t_1+u_1$, with $m_t$ the top quark mass.
For this process, which has complex color flow,  
$H$, $S$, and $\Gamma_S$ are $2\times 2$ matrices.
However, $H^{(0)}$ has a particularly simple form in a singlet-octet 
color basis (the only non-zero element is $H_{22}^{(0)}$ \cite{NKtop,KLMV})
so that simplifications arise.

The NLO coefficients of Section 3 here take the values
$c_3=4C_F$ and $c_2=c_2^{\mu}+T_2$, with 
$c_2^{\mu}=-2C_F \ln(\mu_F^2/m_t^2)$ 
and
\beq
T_2=2\, {\rm Re} \Gamma_{S,22}^{(1)}
-2C_F-2C_F\ln\left(\frac{t_1u_1}{m_t^4}\right)
-2C_F \ln\left(\frac{m_t^2}{s}\right) \, ,
\eeq
where ${\rm Re} \Gamma_{S,22}^{(1)}$ denotes the real part of 
the 22 element of the soft anomalous dimension matrix \cite{KS,NKtop},
\beq
\Gamma_{S,22}^{(1)}=C_F\left[4\ln\left(\frac{u_1}{t_1}\right)
-L_{\beta} -i \pi \right]+\frac{C_A}{2}
\left[-3\ln\left(\frac{u_1}{t_1}\right)
-\ln\left(\frac{m_t^2 s}{t_1u_1}\right)+L_{\beta} +i \pi\right]
\eeq
with
$L_{\beta}=[(1-2m_t^2/s)/\beta][\ln((1-\beta)/(1+\beta))+i \pi]$
and $\beta=\sqrt{1-4m_t^2/s}$.
Note that even though we have nontrivial color matrices for this process, 
due to the simple form of the $H$ matrix here 
we have aborbed the term $2 {\rm Re} \Gamma_{S,22}^{(1)}$, which arises
from $A^c$, into $T_2$. To be precise, 
$A^c=(\sigma^B/ \alpha_s^2) \, 2\, {\rm Re} \Gamma_{S,22}^{(1)}$. 
Also, $c_1=c_1^{\mu}+T_1$ with 
\beqa
c_1^{\mu}&=&C_F \left[\ln\left(\frac{t_1u_1}{m_t^4}\right)-\frac{3}{2}\right]
\ln\left(\frac{\mu_F^2}{m_t^2}\right)
+\frac{\beta_0}{2}\ln\left(\frac{\mu_R^2}{m_t^2}\right)\, 
\eeqa
and $T_1$ as given by the $\delta(s_4)$ terms in Eq. (4.7) of 
Ref. \cite{NLOqqtop} (we have effectively
absorbed $T_1^c$ into $c_1$; for details see \cite{NKRVtop}).
Also,
$F^c=(\sigma^B/ \alpha_s^2)
[4 ({\rm Re} \Gamma_{S,22}^{(1)})^2+4 \Gamma_{S,12}^{(1)}
\Gamma_{S,21}^{(1)}]$ with $\Gamma_{S,12}^{(1)}=(C_F/C_A)
\ln(u_1/t_1)$ and $\Gamma_{S,21}^{(1)}=2\ln(u_1/t_1)$ the 12 and 21
elements of the soft anomalous dimension matrix.

\begin{figure}
\begin{center}
\includegraphics[width=12.5cm]{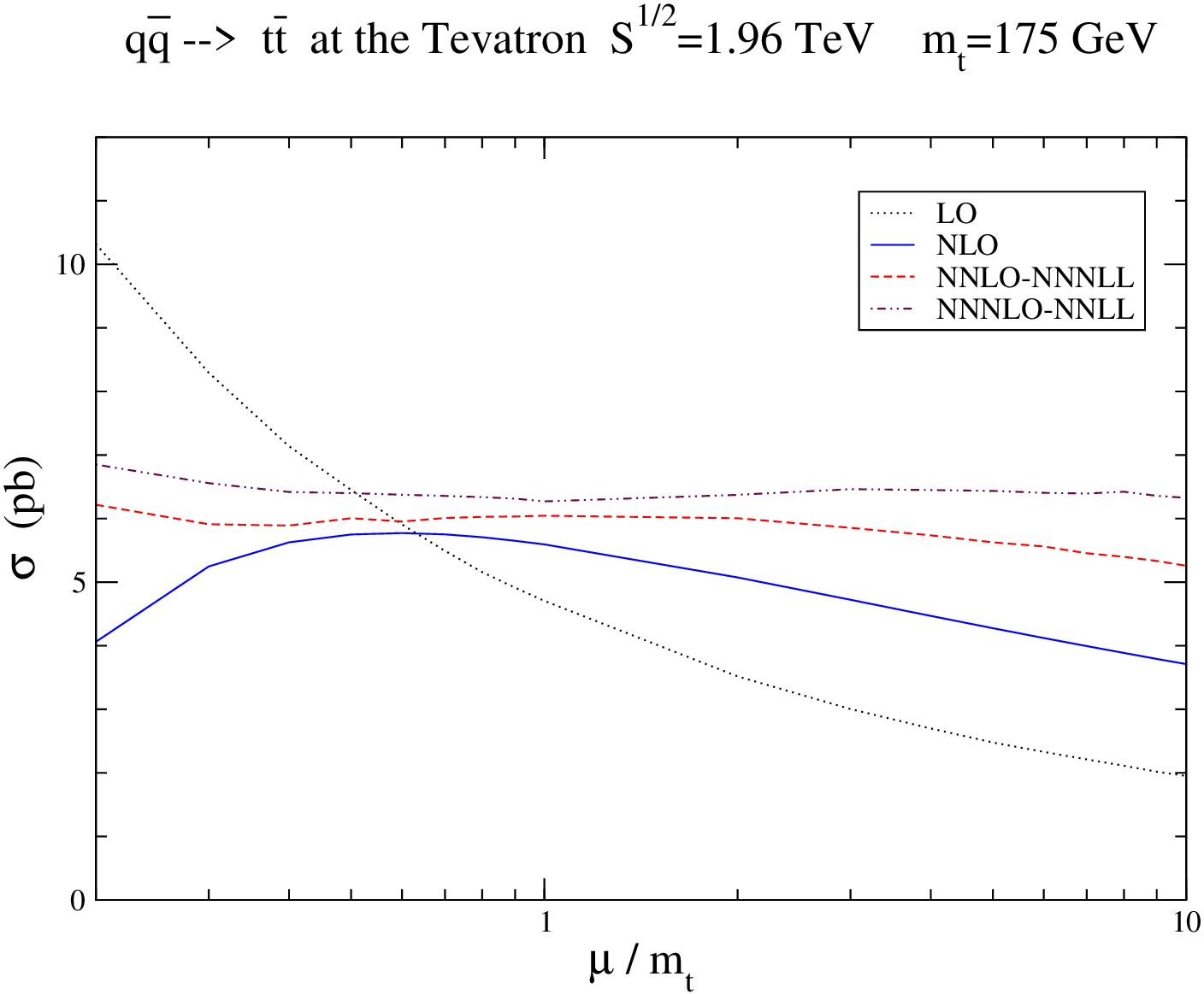}
\caption{The scale dependence of top production 
in the $q {\bar q}$ channel at the Tevatron.}
\label{topqqmuplot}
\end{center}
\end{figure}

The various terms from the Appendix used in the NNNLO corrections 
here take the values  
$X_3=\beta_0 \, c_3/12$, 
$X_2=-(\beta_0/4)T_2+(\beta_0/8)c_3 \ln(\mu_R^2/m_t^2)
+c_3\, K/4$,
$X_2^{\mu}=(\beta_0/8)c_3\ln(\mu_R^2/m_t^2)$,
\beq
X_1^{\mu^2}=-\frac{\beta_0}{4}c_2^{\mu}
\ln\left(\frac{\mu_R^2}{m_t^2}\right)
-C_F\frac{\beta_0}{4}\ln^2\left(\frac{\mu_F^2}{m_t^2}\right)\, ,
\eeq
\beq
X_1^{\mu}=-\frac{\beta_0}{4}\, T_2\, \ln\left(\frac{\mu_R^2}{m_t^2}\right)
+C_F\, K\, \ln\left(\frac{\mu_F^2}{m_t^2}\right) \, ,
\eeq
\beq
X_1^{\zeta}=\frac{\beta_0}{4}\, \zeta_2\, c_3 \, ,
\eeq
\beqa
X_0^{\mu^2}&=&\frac{\beta_0}{4} c_1^{\mu}
\ln\left(\frac{\mu_R^2}{m_t^2}\right)
-\frac{\beta_0^2}{16}\ln^2\left(\frac{\mu_R^2}{m_t^2}\right)
\nonumber \\ && \hspace{-10mm}
{}+\frac{\beta_0}{8}\left[\frac{3}{2}C_F
-C_F\ln\left(\frac{t_1 u_1}{m_t^4}\right)\right]
\ln^2\left(\frac{\mu_F^2}{m_t^2}\right)\, ,
\eeqa
\beq
X_0^{\zeta}=\frac{\beta_0}{6}\zeta_3 c_3
-\frac{\beta_0}{4}\zeta_2 T_2
+\frac{\beta_0}{8}\zeta_2 c_3 \ln\left(\frac{\mu_R^2}{m_t^2}\right)
+C_F \zeta_2 K \, ,
\eeq
\beq
Y_q^{(1,\mu^2)}=C_F \, \frac{\beta_0^2}{8} \,
\ln^2\left(\frac{\mu_F^2}{\mu_R^2}\right) \, ,
\eeq
\beq
Y_q^{(0,\mu^3)}=-C_F \, \frac{\beta_0^2}{48} \,
\left[\ln^3\left(\frac{\mu_F^2}{\mu_R^2}\right) 
+\ln^3\left(\frac{\mu_R^2}{m_t^2}\right) \right]\, ,
\eeq
where $X_i^{\mu^n}$ denotes the terms with scale logarithms to the $n$-th 
power in $X_i$, $X_0^{\zeta}$ denotes the $\zeta_i$ terms in $X_0$,
$Y_q^{(1,\mu^2)}$ denotes terms involving squares of the scale logarithms 
in $Y_q^{(1)}$,
and $Y_q^{(0,\mu^3)}$ denotes terms involving cubes of the scale logarithms 
in $Y_q^{(0)}$.

The NNNLO-NNLL corrections are then given by 
\beqa
{\hat{\sigma}}^{(3)}&=& 
\sigma^B \frac{\alpha_s^3(\mu_R^2)}{\pi^3} \;  
\left \{\frac{1}{8} \, c_3^3 \; {\cal D}_5(s_4)
+\left[\frac{5}{8} \, c_3^2 \, c_2 -\frac{5}{2} \, c_3 \, X_3\right] \;  
{\cal D}_4(s_4) \right.
\nonumber \\ && \hspace{-10mm}
{}+\left[c_3 \, c_2^2 +\frac{1}{2} \, c_3^2 \, c_1
-\zeta_2 \, c_3^3 +(\beta_0-4\, c_2) X_3 +2 \, c_3 \, X_2
+4\, c_3\, \Gamma_{S,12}^{(1)} \, \Gamma_{S,21}^{(1)}
\right] \; {\cal D}_3(s_4)
\nonumber \\ && \hspace{-10mm} 
{}+\left[\frac{3}{2}\, c_3 \,c_2^{\mu} \, c_1 
+\frac{3}{2}\, c_3 \,T_2 \, c_1^{\mu}
+\frac{1}{2} \, \left(c_2^3- \, T_2^3\right)
-3\, \zeta_2 \, c_3^2 \,c_2 +\frac{5}{2} \, \zeta_3 \, c_3^3 
-3\, c_1^{\mu} \, X_3+\frac{27}{2} \, \zeta_2 \, c_3 \, X_3 \right.
\nonumber \\ && \hspace{-5mm} \left.
{}+(3\, T_2-\beta_0) \, X_2^{\mu}+3\, c_2^{\mu} X_2 
-\frac{3}{2}\, c_3 \, \left(X_1^{\mu}+X_1^{\mu^2}+X_1^{\zeta}\right)
+6\,c_2^{\mu}\, \Gamma_{S,12}^{(1)} \, \Gamma_{S,21}^{(1)} \right] 
\; {\cal D}_2(s_4)  
\nonumber \\ &&  \hspace{-10mm}
{}+\left[\frac{1}{2}c_3 \, (c_1^{\mu})^2+(c_2^{\mu})^2 \, c_1
+2\, T_2 \, c_2^{\mu}\, c_1^{\mu}
-\zeta_2 \, c_3^2 \, c_1-\frac{5}{2}\, \zeta_2 \, c_3 \, c_2^2
+5 \, \zeta_3 \, c_3^2 \, c_2 +\frac{5}{4}\, \zeta_2^2 \, c_3^3 
-\frac{15}{4}\, \zeta_4 \, c_3^3 \right.
\nonumber \\ && \hspace{-5mm} 
{}-\frac{\beta_0^2}{4}\,\zeta_2\, c_3
+(-20\, \zeta_3 \, c_3+12 \, \zeta_2 \, c_2) \, X_3 
+2\, c_1^{\mu} \, X_2^{\mu}-5\, \zeta_2 \, c_3 \, X_2 
+(\beta_0-2\,c_2)\left(X_1^{\mu^2}+X_1^{\zeta}\right)
\nonumber \\ && \hspace{-5mm} \left.
{}- 2 \, c_2^{\mu} \, X_1^{\mu}+c_3 \left(X_0^{\mu^2}+X_0^{\zeta}\right)
+2\,Y_q^{(1,\mu^2)}-10\, \zeta_2 \, c_3 \, \Gamma_{S,12}^{(1)} \, 
\Gamma_{S,21}^{(1)}\right] \; {\cal D}_1(s_4)
\nonumber \\ && \hspace{-10mm}
{}+\left[\frac{1}{2}\, c_2^{\mu} \, (c_1^{\mu})^2
-\frac{15}{4} \, \zeta_4 \, c_3^2 \, c_2^{\mu}
+\zeta_3 \, c_3^2 \, c_1^{\mu}
+ 2 \, \zeta_3 \, c_3 \, \left(c_2^2-T_2^2\right) 
+\frac{5}{4} \, \zeta_2^2 \, c_3^2 \, c_2^{\mu}
-\zeta_2 \, c_3 \, c_2 \, c_1^{\mu} \right.
\nonumber \\ && \hspace{-5mm}
{}-\zeta_2 \, c_3 \, c_2^{\mu} \, T_1
-\frac{1}{2} \, \zeta_2 \, \left(c_2^3-T_2^3\right)
+(-8 \, \zeta_3 \, c_2^{\mu} +3 \, \zeta_2 \, c_1^{\mu}) \, X_3
+(4 \, \zeta_3 \, c_3 -3 \, \zeta_2 \, T_2) \, X_2^{\mu}
\nonumber \\ && \hspace{-5mm}
{}-3 \, \zeta_2 \, c_2^{\mu} \, X_2
+\zeta_2 \, c_3\, \left(X_1^{\mu}+X_1^{\mu^2}\right)
-c_1^{\mu} \, \left(X_1^{\mu^2}+X_1^{\zeta}\right)
+c_2^{\mu} \, \left(X_0^{\mu^2}+X_0^{\zeta}\right)
+2\, Y_q^{(0,\mu^3)} 
\nonumber \\ && \hspace{-5mm} \left. \left.
{}-6\, \zeta_2 \, c_2^{\mu} \, \Gamma_{S,12}^{(1)} \, 
\Gamma_{S,21}^{(1)}\right] \; {\cal D}_0(s_4) \right\} \, .
\nonumber \\
\label{NNNLOtop}
\eeqa

Because we have absorbed $A^c$ into $c_2$, and $T_1^c$ into $c_1$,
the corrections take a simple form, simply multiplying the LO term $\sigma^B$.
Note that consistent with a NNLL calculation we include all
${\cal D}_5$, ${\cal D}_4$, and ${\cal D}_3$ terms. In addition, we calculate 
all scale logarithms at NNLL accuracy. 
This means that for coefficients of 
$\ln^i(\mu^2/m^2)$ we include the most singular plus distribution
and the two next-most-singular ones \cite{KLMV}. So we also include
all scale logarithms in the ${\cal D}_2$ terms, the cubed and 
squared scale logarithms in the ${\cal D}_1$ terms, and the cubed 
scale logarithms in the ${\cal D}_0$ terms.  
With respect to the subleading $\zeta_i$ terms that arise from inversion from 
moment to momentum space, we include only those that we can calculate 
exactly (see Ref. \cite{NKtop}). 
Thus we include all $\zeta_i$ terms
in the ${\cal D}_2$ and ${\cal D}_1$ terms, and all $\zeta_i$ terms 
multiplying scale logarithms in the ${\cal D}_0$ term.

In Fig. \ref{topqqmuplot} we plot the scale dependence of the
cross section for $q {\bar q} \rightarrow t {\bar t}$ 
at the Tevatron Run II for a top quark mass $m_t=175$ GeV.
We set the factorization scale equal to the renormalization scale
and denote this common scale by $\mu$.
Again we use the MRST2002 approximate NNLO parton distribution 
functions \cite{MRST2002} (results using the CTEQ6M distributions 
\cite{CTEQ6M}
are similar \cite{NKRVtop}).
We plot a large range in scale and see 
that the higher-order soft-gluon corrections greatly decrease the scale 
dependence of the cross section. The NNLO-NNNLL and the NNNLO-NNLL curves 
are relatively flat. The NNNLO-NNLL result displays very little variation 
and it approaches the scale independence 
expected of a physical cross section.

\mysection{Conclusions}

In this paper, I presented a unified approach to calculating the
NNNLO soft-gluon corrections for hard-scattering processes in hadron-hadron
and lepton-hadron collisions in the $\overline{\rm MS}$ scheme 
in either 1PI or PIM kinematics. 
The master formulas given in the paper allow explicit calculations for
any process, with either simple or complex color flows,
keeping in general the factorization and renormalization scales
separate and the color factors explicit. 
Detailed results, illustrating the use of the master formulas, 
were given to NLL accuracy for charged Higgs production via bottom-gluon 
fusion at the LHC, and to NNLL accuracy for top quark production via 
quark-antiquark annihilation at the Tevatron.

The NNLO and NNNLO corrections increase theoretical accuracy and diminish 
the dependence on the factorization and renormalization scales, and thus 
are essential in further calculations of QCD corrections for many 
hard-scattering processes, both in their own right and as backgrounds which 
may be particularly important in searching for the Higgs boson and 
supersymmetric particles, 
as well as other processes that signal new physics beyond the Standard Model.

\renewcommand{\theequation}{A.\arabic{equation}}
\setcounter{equation}{0}  
\section*{Appendix}

The $\beta$ function is given by
\beq
\beta(\alpha_s) \equiv \mu \, d \ln g/d \mu
=-\beta_0 \alpha_s/(4 \pi)-\beta_1 \alpha_s^2/(4 \pi)^2 
-\beta_2 \alpha_s^3/(4 \pi)^3+\cdots \, ,
\eeq
where $g^2=4 \pi \alpha_s$, 
with $\beta_0=(11C_A-2n_f)/3$,  $\beta_1=34 C_A^2/3-2n_f(C_F+5C_A/3)$,
and 
\beq
\beta_2=\frac{2857}{54}C_A^3+\left(C_F^2-\frac{615}{54}C_F C_A
-\frac{1415}{54}C_A^2\right)n_f
+\left(\frac{33}{27}C_F+\frac{79}{54}C_A\right)n_f^2 \, .
\eeq
Note that 
\beqa
\alpha_s(\mu)&=&\alpha_s(\mu_R)\left[1-\frac{\beta_0}{4\pi}\alpha_s(\mu_R)
\ln\left(\frac{{\mu}^2}{\mu_R^2}\right)
+\frac{\beta_0^2}{16\pi^2}\alpha_s^2(\mu_R)
\ln^2\left(\frac{{\mu}^2}{\mu_R^2}\right)
-\frac{\beta_1}{16\pi^2}\alpha_s^2(\mu_R)
\ln\left(\frac{{\mu}^2}{\mu_R^2}\right) \right.
\nonumber \\ && \hspace{20mm} \left.
+\cdots\right] \, .
\eeqa

The various quantities used in the NNNLO expression, Eq. (\ref{NNNLOmaster}),
are given by the following expressions.
For the quantities $X_3$, $X_2$, $X_1$, $X_0$, we have
\beq
X_3=\frac{\beta_0}{12} c_3-\sum_j C_j\frac{\beta_0}{24} \, ,
\eeq
\beq
X_2=-\frac{\beta_0}{4}T_2+\frac{\beta_0}{8}c_3 \ln\left(\frac{\mu_R^2}{M^2}
\right)+c_3\frac{K}{4}-\sum_j\frac{\beta_0}{8}B_j^{(1)}\, ,
\eeq
\beqa
X_1&=&c_2\, c_1-\zeta_2\, c_3\, c_2+\zeta_3 \, c_3^2 
+\frac{\beta_0}{4}\, \zeta_2 \, c_3 -\sum_j C_j \frac{\beta_0}{8} \zeta_2
-C_{D_0}^{(2)} \, , 
\eeqa
and
\beqa
X_0&=&\frac{\beta_0}{4} c_1 \ln\left(\frac{\mu_R^2}{M^2}\right)
+\frac{d_{\alpha_s}}{16}\left[-\frac{\beta_0^2}{2}\ln^2\left(\frac{\mu_R^2}
{M^2}\right)
-\beta_0^2\, \ln\left(\frac{\mu_R^2}{M^2}\right)
\ln\left(\frac{M^2}{s}\right)
+\beta_1\ln\left(\frac{\mu_R^2}{M^2}\right)\right]
\nonumber \\ && \hspace{-5mm}
{}+\frac{\beta_0}{6}\zeta_3 c_3-\frac{\beta_0}{4}\zeta_2 T_2
+\frac{\beta_0}{8} \zeta_2 c_3 \ln\left(\frac{\mu_R^2}{M^2}\right)
+\sum_i C_i \frac{\zeta_2}{2} K
-\sum_i {\gamma'}_{i/i}^{(2)} \ln\left(\frac{\mu_F^2}{M^2}\right)
\nonumber \\ && \hspace{-5mm}
{}+\sum_i \frac{\beta_0}{8}\left[\gamma_i^{(1)}-C_i
\ln\left(\frac{-t_i}{M^2}\right)\right] 
\left[\ln^2\left(\frac{\mu_F^2}{M^2}\right)
+2\, \ln\left(\frac{\mu_F^2}{M^2}\right)\ln\left(\frac{M^2}{s}\right)\right]
\nonumber \\ && \hspace{-5mm}
{}+\sum_i C_i \frac{K}{2} \ln\left(\frac{-t_i}{M^2}\right)
\ln\left(\frac{\mu_F^2}{M^2}\right)
-\sum_j C_j \frac{\beta_0}{12}\zeta_3 
-\sum_j C_j \zeta_2 \frac{K}{4}-\sum_j \frac{\beta_0}{8} 
\zeta_2 B_j^{(1)} +{R'}^{(2)}.
\eeqa
The quantity ${R'}^{(2)}$ at the end of the above equation stands for the
virtual two-loop corrections $R^{(2)}$ in Eq. (\ref{NNLOmaster}) 
minus $\zeta$ terms and scale terms.
To be precise
\beqa
R^{(2)}&=&{R'}^{(2)}+\frac{1}{2} c_1^2
-\frac{\zeta_2}{2}\, c_2^2
+\frac{1}{4}\zeta_2^2 \, c_3^2
+\zeta_3 \, c_3 \, c_2-\frac{3}{4}\zeta_4 \, c_3^2
+\frac{\beta_0}{4} c_1 \ln\left(\frac{\mu_R^2}{M^2}\right)
\nonumber \\ &&  
{}+\frac{d_{\alpha_s}}{16}\left[-\frac{\beta_0^2}{2}
\ln^2\left(\frac{\mu_R^2}{M^2}\right)
-\beta_0^2\, \ln\left(\frac{\mu_R^2}{M^2}\right)
\ln\left(\frac{M^2}{s}\right)
+\beta_1\ln\left(\frac{\mu_R^2}{M^2}\right)\right]
\nonumber \\ &&
{}-\sum_i {\gamma'}_{i/i}^{(2)} \ln\left(\frac{\mu_F^2}{M^2}\right)
+\sum_i C_{f_i} \frac{K}{2} \, \ln\left(\frac{-t_i}{M^2}\right)
\ln\left(\frac{\mu_F^2}{M^2}\right)
\nonumber \\ && 
{}+\sum_i \frac{\beta_0}{8} \left[\gamma_i^{(1)}
-C_{f_i} \, \ln\left(\frac{-t_i}{M^2}\right)\right]
\left[\ln^2\left(\frac{\mu_F^2}{M^2}\right)
+2\, \ln\left(\frac{\mu_F^2}{M^2}\right)
\ln\left(\frac{M^2}{s}\right) \right] \, .
\eeqa

Also
\beqa
Y_i^{(1)}&=&2 A_i^{(3)}
+C_i \beta_0 \left[\frac{\beta_0}{8}
\ln^2\left(\frac{\mu_F^2}{\mu_R^2}\right)
-\frac{K}{2} \ln\left(\frac{\mu_F^2}{\mu_R^2}\right)\right]
\nonumber \\ &&
{}+C_i \frac{\beta_1}{8} \left[\ln\left(\frac{\mu_R^2}{M^2}\right)
+\ln\left(\frac{M^2}{s}\right)
+1+2\ln\left(\frac{-t_i}{M^2}\right)\right] \, ,
\nonumber \\
\eeqa
\beqa
Y_j^{(1)}&=&{}-\frac{\beta_0}{2}B_j^{(2)}-A_j^{(3)}
-C_j \frac{\beta_0^2}{16} \ln^2\left(\frac{\mu_R^2}{M^2}\right)
+\frac{\beta_0}{8}\left[-\beta_0 B_j^{(1)}-2 C_j K\right]
\ln\left(\frac{\mu_R^2}{M^2}\right)
\nonumber \\ && \hspace{-5mm}
{}+\frac{C_j}{8}\zeta_2 \beta_0^2  
+\frac{\beta_1}{16}\left[-C_j \ln\left(\frac{\mu_R^2}{M^2}\right)
+2C_j+B_j^{(1)}+2C_j \ln\left(\frac{M^2}{s}\right)\right] \, ,
\eeqa
\beqa
Y_i^{(0)}&=&
{}-C_i \frac{\beta_0^2}{48}
\left[\ln^3\left(\frac{\mu_F^2}{\mu_R^2}\right)
+\ln^3\left(\frac{\mu_R^2}{M^2}\right)\right]
+\frac{C_i}{32}\left(\beta_1+4\beta_0 K\right) 
\left[\ln^2\left(\frac{\mu_F^2}{\mu_R^2}\right)
-\ln^2\left(\frac{\mu_R^2}{M^2}\right)\right]
\nonumber \\ && \hspace{-5mm}
{}-C_i \frac{\beta_0}{4} K \left[2\ln\left(\frac{-t_i}{M^2}\right)
+\ln\left(\frac{M^2}{s}\right)\right]
\ln\left(\frac{\mu_R^2}{M^2}\right)
-C_i \frac{\beta_1}{16} \left[1+2\ln\left(\frac{-t_i}{M^2}\right)\right]
\ln\left(\frac{\mu_R^2}{M^2}\right) 
\nonumber \\ && \hspace{-5mm}
{}-\frac{\beta_0^2}{16} \left[4\gamma_i^{(1)}+2C_i-\beta_0\, d_{\alpha_s}
+C_i\frac{\beta_1}{\beta_0^2}+C_i\ln\left(\frac{M^2}{s}\right)\right]
\ln\left(\frac{M^2}{s}\right) \ln\left(\frac{\mu_R^2}{M^2}\right)
\nonumber \\ && \hspace{-5mm}
{}-{\nu}_i^{(2)} \frac{\beta_0}{2} \ln\left(\frac{\mu_R^2}{M^2}\right)
-A_i^{(3)} \ln\left(\frac{\mu_F^2}{M^2}\right)-{\nu}_i^{(3)}
-\frac{C_i}{6} \beta_0^2 \ln^3\left(\frac{-t_i}{M^2}\right) 
-\frac{C_i}{4} \beta_0 (\beta_0+2K) 
\ln^2\left(\frac{-t_i}{M^2}\right)
\nonumber \\ && \hspace{-5mm}
{}-\left(\beta_0 {\nu}_i^{(2)}+2 A_i^{(3)}\right) 
\ln\left(\frac{-t_i}{M^2}\right)
-\frac{\beta_0^2}{48}C_i \ln^3\left(\frac{M^2}{s}\right)
-\frac{C_i}{32}\left(\beta_1+4\beta_0 K\right)\ln^2\left(\frac{M^2}{s}\right)
\nonumber \\ && \hspace{-5mm}
{}+\frac{\beta_0^2}{4}\left[\frac{C_i}{2}\ln\left(\frac{-t_i}{M^2}\right)
-\gamma_i^{(1)}-\frac{C_i}{4}+\frac{\beta_0}{4}d_{\alpha_s}\right]
\ln^2\left(\frac{M^2}{s}\right)
\nonumber \\ && \hspace{-5mm}
{}+\left[-A_i^{(3)}-\nu_i^{(2)}\frac{\beta_0}{2}
-C_i K\frac{\beta_0}{2}\ln\left(\frac{-t_i}{M^2}\right)
+\frac{\beta_1 \beta_0}{32}d_{\alpha_s}-C_i\frac{\beta_1}{16}
-\frac{\beta_1}{8}\gamma_i^{(1)}\right]\ln\left(\frac{M^2}{s}\right) \, ,
\eeqa
and
\beqa
Y_j^{(0)}&=&{}\frac{\beta_0}{4}\left[-2 B_j^{(2)}-2{\nu}_j^{(2)}
-\left(\frac{\beta_0}{2}(2B_j^{(1)}+C_j)+C_j K\right) 
\ln\left(\frac{M^2}{s}\right)
-C_j \frac{\beta_0}{4}\ln^2\left(\frac{M^2}{s}\right)\right]
\ln\left(\frac{\mu_R^2}{M^2}\right)
\nonumber \\ && \hspace{-5mm}
{}-\frac{\beta_1}{16} \left[B_j^{(1)}+C_j
+C_j \ln\left(\frac{M^2}{s}\right)\right]
\ln\left(\frac{\mu_R^2}{M^2}\right)-B_j^{(3)}-{\nu}_j^{(3)}
-\frac{7}{48} C_j \beta_0^2 \ln^3\left(\frac{M^2}{s}\right)
\nonumber \\ && \hspace{-5mm}
{}+\frac{\beta_0}{16}\left(-5C_j\beta_0-4\beta_0 B_j^{(1)}
+2 C_j K\right) \ln^2\left(\frac{M^2}{s}\right)
-\frac{\beta_1}{32} C_j \ln^2\left(\frac{M^2}{s}\right)
\nonumber \\ && \hspace{-5mm}
{}+\left[\frac{\beta_0}{2} {\nu}_j^{(2)}
-A_j^{(3)}-\frac{\beta_1}{16}
\left(2 B_j^{(1)}+C_j\right)\right] \ln\left(\frac{M^2}{s}\right) \, .
\eeqa

Also we have defined,
\beqa
M^c&=&{\rm tr}  \left[H^{(1)} \Gamma_S^{(1)\,\dagger} S^{(0)}
+H^{(1)} S^{(0)} \Gamma_S^{(1)} + H^{(0)} \Gamma_S^{(1)\,\dagger} S^{(1)}
+H^{(0)} S^{(1)} \Gamma_S^{(1)}\right] \, ,
\eeqa
\beqa
K_1^c&=&{\rm tr} \left[H^{(2)} \Gamma_S^{(1)\, \dagger} S^{(0)}
+H^{(0)} S^{(2)} \Gamma_S^{(1)} 
+H^{(2)} S^{(0)} \Gamma_S^{(1)} \right. 
\nonumber \\ &&
{}+H^{(0)} \Gamma_S^{(1)\, \dagger} S^{(2)}
+H^{(1)} \Gamma_S^{(1)\, \dagger} S^{(1)}
+H^{(1)} S^{(1)} \Gamma_S^{(1)} 
\nonumber \\ &&
{}+H^{(1)} \Gamma_S^{(2)\, \dagger} S^{(0)}
+H^{(1)} S^{(0)} \Gamma_S^{(2)} 
+H^{(0)} \Gamma_S^{(2)\, \dagger} S^{(1)}
\nonumber \\ && \left.
{}+H^{(0)} S^{(1)} \Gamma_S^{(2)} 
+H^{(0)} \Gamma_S^{(3)\, \dagger} S^{(0)}
+H^{(0)} S^{(0)} \Gamma_S^{(3)} \right] \, ,
\eeqa
\beqa
K_2^c&=&{\rm tr} \left[
H^{(1)} \left(\Gamma_S^{(1)\, \dagger}\right)^2 S^{(0)}
+H^{(0)} \left(\Gamma_S^{(1)\, \dagger}\right)^2 S^{(1)}
+\left(H^{(1)} S^{(0)}+H^{(0)} S^{(1)}\right) 
\left(\Gamma_S^{(1)}\right)^2 \right.
\nonumber \\ &&
{}+2 \, H^{(1)} \Gamma_S^{(1)\, \dagger} S^{(0)} \Gamma_S^{(1)}
+2 \, H^{(0)} \Gamma_S^{(1)\, \dagger} S^{(1)} \Gamma_S^{(1)}
+2  \, H^{(0)} \Gamma_S^{(1)\, \dagger} \Gamma_S^{(2)\, \dagger} S^{(0)}
\nonumber \\ && \left.
{}+2 \,  H^{(0)} S^{(0)} \Gamma_S^{(1)} \Gamma_S^{(2)}
+2\, H^{(0)} \Gamma_S^{(2)\, \dagger} S^{(0)} \Gamma_S^{(1)}
+2\, H^{(0)} \Gamma_S^{(1)\, \dagger} S^{(0)} \Gamma_S^{(2)}\right] \, ,
\eeqa
and
\beqa
K_3^c&=&{\rm tr} \left[
H^{(0)} \left(\Gamma_S^{(1)\, \dagger}\right)^3 S^{(0)}
+H^{(0)} S^{(0)} \left(\Gamma_S^{(1)}\right)^3 \right. 
\nonumber \\ && \left.
{}+3\, H^{(0)} \left(\Gamma_S^{(1)\, \dagger}\right)^2
S^{(0)} \Gamma_S^{(1)}+ 3\, H^{(0)} \Gamma_S^{(1)\, \dagger}
S^{(0)}  \left(\Gamma_S^{(1)}\right)^2 \right] \, .
\eeqa


\begin{thebibliography}{99}

\bibitem{QCDSM}
The QCD/SM Working Group: Summary Report, 
in {\sl Les Houches 2003, Physics at TeV colliders}, p. 291,
hep-ph/0403100, and references therein.

\bibitem{GS}
G. Sterman, Nucl. Phys. B {\bf 281}, 310 (1987).

\bibitem{CT}
S. Catani and L. Trentadue, Nucl. Phys. B {\bf 327}, 323 (1989).

\bibitem{CLS}
H. Contopanagos, E. Laenen, and G. Sterman,  
Nucl. Phys. B {\bf 484}, 303 (1997) [hep-ph/9604313].

\bibitem{KS}
N. Kidonakis and G. Sterman, Phys. Lett. B {\bf 387}, 867 (1996);
Nucl. Phys. B {\bf 505}, 321 (1997) [hep-ph/9705234].

\bibitem{KOS}
N. Kidonakis, G. Oderda, and G. Sterman,
Nucl. Phys. B {\bf 525}, 299 (1998) [hep-ph/9801268];
Nucl. Phys. B {\bf 531}, 365 (1998) [hep-ph/9803241].

\bibitem{LOS}
E. Laenen, G. Oderda, and G. Sterman, 
Phys. Lett. B {\bf 438}, 173 (1998) [hep-ph/9806467].

\bibitem{NK3nlo}
N. Kidonakis, in {\sl DPF 2004}, Int. J. Mod. Phys. A {\bf 20}, 3726 (2005)
[hep-ph/0410116];
in {\sl DIS 2005}, hep-ph/0506299.

\bibitem{NKtop} 
N. Kidonakis, Phys. Rev. D {\bf 64}, 014009 (2001) [hep-ph/0010002].

\bibitem{NKmpla}
N. Kidonakis, Mod. Phys. Lett. A {\bf 19}, 405 (2004) [hep-ph/0401147], 
and references therein.

\bibitem{GKS}
N. Kidonakis and A. Sabio Vera, JHEP {\bf 02}, 027 (2004) 
[hep-ph/0311266]; 
R.J. Gonsalves, N. Kidonakis, and A. Sabio Vera, hep-ph/0507317;
N. Kidonakis and V. Del Duca, Phys. Lett. B {\bf 480}, 87 (2000) 
[hep-ph/9911460]. 

\bibitem{KOp}
N. Kidonakis and J.F. Owens, Phys. Rev. D {\bf 61}, 094004 (2000) 
[hep-ph/9912388];
Int. J. Mod. Phys. A {\bf 19}, 149 (2004) [hep-ph/0307352].

\bibitem{KLMV} N. Kidonakis, E. Laenen, S. Moch, and R. Vogt, 
Phys. Rev. D {\bf 64}, 114001 (2001) [hep-ph/0105041];
N. Kidonakis, in {\sl DPF 2000}, Int. J. Mod. Phys. A {\bf 16}, 
s1A, 363 (2001) [hep-ph/0009013].

\bibitem{NKRVtop}
N. Kidonakis and R. Vogt, Phys. Rev. D {\bf 68}, 114014 (2003) 
[hep-ph/0308222];
Eur. Phys. J. C {\bf 33}, s466 (2004) [hep-ph/0309045];
in {\sl DPF 2004}, Int. J. Mod. Phys. A {\bf 20}, 3171 (2005) [hep-ph/0410367].

\bibitem{KVbc}
N. Kidonakis, E. Laenen, S. Moch, and R. Vogt, 
Phys. Rev. D {\bf 67}, 074037 (2003) [hep-ph/0212173];
Nucl. Phys. A {\bf 715}, 549 (2003) [hep-ph/0208119];
N. Kidonakis and R. Vogt, Eur. Phys. J. C {\bf 36}, 201 (2004) 
[hep-ph/0401056].

\bibitem{LM}
E. Laenen and S. Moch, Phys. Rev. D {\bf 59}, 034027 (1999) [hep-ph/9809550].

\bibitem{NYI}
N.Y. Ivanov, Nucl. Phys. B {\bf 615}, 266 (2001) [hep-ph/0104301].

\bibitem{NKchiggs}
N. Kidonakis, JHEP {\bf 05}, 011 (2005) [hep-ph/0412422];
in {\sl DIS 2005}, hep-ph/0505271.

\bibitem{KOj}
N. Kidonakis and J.F. Owens, Phys. Rev. D {\bf 63}, 054019 (2001) 
[hep-ph/0007268].

\bibitem{SUSY}
Beyond the Standard Model Working Group:  Summary Report, 
in {\sl Les Houches 2003, Physics at TeV colliders}, p. 171,
hep-ph/0402295, and references therein. 

\bibitem{Higgs}
The Higgs Working Group: Summary Report, 
in {\sl Les Houches 2003, Physics at TeV colliders}, p. 1,
hep-ph/0406152, and references therein. 

\bibitem{BK}
A. Belyaev and N. Kidonakis, Phys. Rev. D {\bf 65}, 037501 (2002) 
[hep-ph/0102072];
N. Kidonakis and A. Belyaev, JHEP {\bf 12}, 004 (2003) [hep-ph/0310299].

\bibitem{NKuni}
N. Kidonakis, Int. J. Mod. Phys. A {\bf 19}, 1793 (2004) [hep-ph/0303186];
in {\sl DIS 2004}, hep-ph/0306125, hep-ph/0307207.

\bibitem{NKrev}
N. Kidonakis, Int. J. Mod. Phys. A {\bf 15}, 1245 (2000) [hep-ph/9902484], 
and references therein.

\bibitem{CB}
C.F. Berger, Phys. Rev. D {\bf 66}, 116002 (2002) [hep-ph/0209107].

\bibitem{MVV}
S. Moch, J.A.M. Vermaseren, and A. Vogt, hep-ph/0506288.

\bibitem{KT}
J. Kodaira and L. Trentadue,  Phys. Lett. B {\bf 112}, 66 (1982). 

\bibitem{BoSt}
J. Botts and G. Sterman, Nucl. Phys. B {\bf 325}, 62 (1989). 

\bibitem{NK2l}
N. Kidonakis, hep-ph/0208056; in {\sl DIS 2003}, hep-ph/0307145.   

\bibitem{BGGS}
A. Belyaev, D. Garcia, J. Guasch, and J. Sola,  
Phys. Rev. D {\bf 65}, 031701 (2002) [hep-ph/0105053];
JHEP {\bf 06}, 059 (2002) [hep-ph/0203031]. 
 
\bibitem{Zhu}
S. Zhu, Phys. Rev. D {\bf 67}, 075006 (2003) [hep-ph/0112109]. 

\bibitem{Plehn}
T. Plehn, Phys. Rev. D {\bf 67}, 014018 (2003) [hep-ph/0206121]. 

\bibitem{BHJP}
E.L. Berger, T. Han, J. Jiang, and T. Plehn, 
Phys. Rev. D {\bf 71}, 115012 (2005) [hep-ph/0312286]. 

\bibitem{MRST2002}
A.D. Martin, R.G. Roberts, W.J. Stirling, and R.S. Thorne,
Eur. Phys. J. C {\bf 28}, 455 (2003) [hep-ph/0211080].

\bibitem{dFV}
D. de Florian and W. Vogelsang, Phys. Rev. D {\bf 71}, 114004 (2005) 
[hep-ph/0501258].

\bibitem{CDF}
CDF Coll., hep-ex/0504053; Phys. Rev. D {\bf 72}, 032002 (2005) 
[hep-ex/0506001].

\bibitem{D0}
D0 Coll., hep-ex/0504043; hep-ex/0504058; 
hep-ex/0505082.

\bibitem{Wagner}
W. Wagner, Rept. Prog. Phys. {\bf 68}, 2409 (2005) [hep-ph/0507207], 
and references therein.

\bibitem{KSpty}
N. Kidonakis and J. Smith, Phys. Rev. D {\bf 51}, 6092 (1995) 
[hep-ph/9502341].   

\bibitem{NLOqqtop}
W. Beenakker, W.L. van Neerven, R. Meng, G.A. Schuler, and J. Smith, 
Nucl. Phys. B {\bf 351}, 507 (1991). 

\bibitem{CTEQ6M}
J. Pumplin, D.R. Stump, J. Huston, H.L. Lai, P. Nadolsky, and W.K. Tung, 
JHEP {\bf 07}, 012 (2002) [hep-ph/0201195].

\end{thebibliography}
\end{document}